\documentclass[reprint,amsmath,amssymb,aps,prd]{revtex4-2}

\usepackage{graphicx}
\usepackage{dcolumn}
\usepackage{bm}
\usepackage{xcolor} 
\usepackage{tabularx} 
\usepackage{amsmath}
\usepackage{ulem}

\newcommand{\comment}[1]{} 

\begin{document}
	
\preprint{APS/123-QED}

\title{Evidence against a strong first-order phase transition in neutron star cores:\\ Impact of new data }

\author{Len Brandes}
\email{len.brandes@tum.de}
\author{Wolfram Weise}%
\email{weise@tum.de}
\author{Norbert Kaiser}%
\email{nkaiser@ph.tum.de}
\affiliation{%
	Technical University of Munich,  TUM School of Natural Sciences,  Physics Department,  85747 Garching, Germany
}%

\date{\today}

\begin{abstract}
	
	With the aim of exploring the evidence for or against phase transitions in cold and dense baryonic matter,  the inference of the sound speed and equation-of-state for dense matter in neutron stars is extended in view of recent new observational data. The impact of the heavy (2.35 $M_\odot$) black widow pulsar PSR J0952-0607 and of the unusually light supernova remnant HESS J1731-347 is inspected.  In addition a detailed re-analysis is performed of the low-density constraint based on chiral effective field theory and of the perturbative QCD constraint at asymptotically high densities,  in order to clarify the influence of these constraints on the inference procedure.  The trace anomaly measure, $\Delta = 1/3 - P/\varepsilon$,  is also computed and discussed. A systematic Bayes factor assessment quantifies the evidence (or non-evidence) of low averaged sound speeds $(c_s^2 \leq 0.1)$, a prerequisite for a phase transition, within the range of densities realized in the core of neutron stars.  One of the consequences of including PSR J0952-0607 in the data base is a further stiffening of the equation-of-state,  resulting for a 2.1 solar-mass neutron star in a reduced central density of less than five times the equilibrium density of normal nuclear matter at the 68\% level. The evidence against small sound speeds in neutron star cores is further strengthened. Within the inferred 68\% posterior credible bands,  only a weak first-order phase transition with a coexistence density interval $\Delta n/n \lesssim 0.2$ would be compatible with the observed data.
	
\end{abstract}

\maketitle

\section{INTRODUCTION}

The inference of the equation-of-state (EoS) using the empirical data base from neutron star observations has progressed significantly in recent times.  Primary sources of information are Shapiro delay measurements of neutron star masses \cite{Demorest2010,Fonseca2016,Arzoumanian2018,Antoniadis2013,Cromartie2020,Fonseca2021},  determinations of masses and radii inferred from x-ray data detected with the NICER telescope \cite{Riley2019,Riley2021,Miller2019,Miller2021,Salmi2022},  and the evaluation of gravitational wave signals from binary neutron star mergers observed by the LIGO Scientific and Virgo Collaborations \cite{Abbott2019,Abbott2020}.  These data have served as standard input sets for a multitude of Bayesian inference analyses \cite{Raaijmakers2021,Pang2021,Brandes2023,Annala2023,Legred2021,Biswas2021,Biswas2022,Takatsy2023,Han2023,Marczenko2023,Ecker2022a,Ecker2022,Jiang2023,Huth2022,Lim2022,Somasundaram2023,Somasundaram2023a,Mroczek2023,Altiparmak2022,Essick2023} in search for a compatible range of EoSs,  pressure $P(\varepsilon)$ as function of energy density $\varepsilon$,  with controlled uncertainties.  An important part of these procedures are the nuclear physics constraints at low densities around the equilibrium density of nuclear matter,  $n_0 = 0.16\,$fm$^{-3}$,  for which chiral effective field theory (ChEFT) results are invoked \cite{Drischler2021,Drischler2022}.  At asymptotically high densities,  perturbative QCD (pQCD) is applicable and provides a constraint that must be matched in extrapolations far beyond the conditions realized in neutron star cores \cite{Gorda2021,Komoltsev2022,Gorda2023}. 

The present work extends our previous Bayesian studies \cite{Brandes2023} in view of new observational data reported in the literature.  We refer primarily to the black widow pulsar PSR J0952-0607 \cite{Romani2022} with a mass $M = 2.35 \pm 0.17\,M_\odot$, in units of the solar mass $M_\odot$,  the heaviest neutron star found so far.  This is also one of the fastest rotating neutron stars.  With a spin period of 1.41 ms (an angular velocity of $4.46\times 10^3$ rotations per second) it requires corrections for rotational effects,  a point that we shall consider in order to incorporate the equivalent non-rotating mass in our data base.

A second recently reported object is the supernova remnant HESS J1731-347 \cite{Doroshenko2022} with an unusually small mass, $M = 0.77^{+0.20}_{-0.17}\,M_{\odot}$,  and radius,  $R = 10.40^{+0.86}_{-0.78}$ km.  The analysis on which these results are based is subject to some discussion \cite{Alford2023} concerning model dependent assumptions about the composition of the remnant's atmosphere.  In fact this object falls well outside the range of masses and radii supported by previous data sets.  It is nonetheless of some interest to examine its impact on the overall systematics of the inference procedure.

Further topics to be investigated in the present paper are the roles of low- and high-density constraints conditioning the inference of the sound speed and EoS.  At low baryon densities up to about $n \sim 2\,n_0$,  ChEFT results have frequently been used as prior input.  We prefer a ChEFT implementation in terms of a likelihood and only up to $n \simeq 1.3\,n_0$,  given that an extension to $2\,n_0$ raises convergence issues \cite{Drischler2021,Drischler2022} as we shall point out.  The pQCD constraint at high densities and its possible impact on the EoS \cite{Gorda2021,Komoltsev2022,Gorda2023,Somasundaram2023a}, even down to densities realized in massive neutron stars,  will also be examined.  We shall find that this impact is quite limited,  given that pQCD methods are restricted to asymptotically large densities above about $40\,n_0$.

A central part of this investigation,  as in previous work,  is a detailed inference of the sound velocity,  $c_s = \sqrt{\partial P(\varepsilon)/\partial\varepsilon}$.  Key questions are whether it exceeds the conformal bound,  $c_s^2 = 1/3$,  at intermediate energy densities,  and the possible approach to conformal matter \cite{Fujimoto2022a} as displayed by a computation of the trace anomaly measure \cite{Marczenko2023},  $\Delta = 1/3 - P/\varepsilon$.  Both items will be systematically examined.  Closely related is a Bayes factor analysis quantifying the likelihood of a rapid change
in energy density over a small range of pressures, which is the
defining characteristic of a first-order phase transition in the core
of heavy neutron stars. Special emphasis is put on the additional impact of PSR J0952-0607 in comparison with the previously employed data.  By first-order phase transition we refer in this work to a `strong` transition subject to a Maxwell construction,  i.e.  with large interface tension between the phases.  

This paper is organised as follows.  Section \ref{sec:Methods} briefly introduces the Bayesian method and the parametrization used to model the speed of sound inside neutron stars.  Sections \ref{sec:BayesianInference} and \ref{sec:likelihoods} explain and summarize the statistical procedures that we use to infer constraints for neutron star properties based on empirical data together with theoretical low-and high-density conditions.  The results for the sound speed and related neutron star properties are presented and discussed in section \ref{sec:Results} where also implications for possible phase changes inside neutron stars are examined.  A summary and conclusions follow in section \ref{sec:Summary}.

\section{Methods}
\label{sec:Methods}

\subsection{Speed of sound}

One starting point for organizing the inference of the equation-of-state of neutron star matter is a general parametrization of the sound speed.  Its square is defined by the derivative of pressure with respect to energy density:
\begin{eqnarray}
	c_s^2(\varepsilon) = \frac{\partial P(\varepsilon)}{\partial \varepsilon}~.
	\label{eq:soundspeed}
\end{eqnarray}
The EoS itself is reconstructed  by simple integration,  $P(\varepsilon) = \int_0^\varepsilon d\varepsilon'\,c_s^2(\varepsilon')$.  Causality and thermodynamic stability dictate that the sound velocity must be within the interval $0 \leq c_s \leq 1$.  Different possible phases of neutron star matter are reflected in the behaviour of $c_s^2$ as a function of energy density \cite{Kojo2021a}. For example, models describing neutron star matter in terms of conventional hadronic (nucleon and meson) degrees of freedom generally display a monotonously rising speed of sound \cite{Akmal1998,Friman2019,Brandes2021}.  On the other hand, the signature of a first-order phase transition (or sufficiently
sharp crossover) would be a rapid change in energy density over a
small range of pressures, manifested in our parametrisation as a
sudden drop of $c_s^2$ at a critical energy density \cite{Benic2015},  while a softer continuous crossover may lead to a peaked behaviour \cite{Baym2018,McLerran2019}.  Some model calculations find that an onset of hyperonic degrees of freedom can lead to a substantial softening of the speed of sound \cite{Motta2021}.  For a survey of possible phases inside neutron stars,  see e.g. Refs.\,\cite{Brandes2023, Fukushima2013}. 

With an equation-of-state $P(\varepsilon)$ deduced from $c_s^2$ as input to the Tolman-Oppenheimer-Volkoff (TOV) equations,  the total mass $M$ and radius $R$ of a neutron star can be computed.  Numerically solving the TOV equations for several central pressures,  $P_c$,  leads to a mass-radius relation,  $M(R)$,  for each given EoS.  This includes a maximum supported mass, $M_{\max}$,  beyond which no stable solution exists.  Additional coupled differential equations can be solved for the tidal deformability $\Lambda$,  relevant for binary neutron star systems \cite{Flanagan2008,Hinderer2008}.

\subsection{Parametrisation}

We prepare a sufficiently general parametrization as an approximation to the squared sound speed,  capable of capturing a wide range of possible phase transitions or crossovers,  in terms of segment-wise linear interpolations.  This ansatz is similar to the one employed in \cite{Annala2020,Annala2022,Altiparmak2022}.  In a previous work \cite{Brandes2023} we systematically compared this parametrization to a speed-of-sound model based on a skewed Gaussian distribution.  It was found that the posterior neutron star properties inferred with these two parametrizations agree within the uncertainties associated with the still limited amount and accuracy of the available astrophysical data.  However,  the segments parametrization turns out to be  preferred as it leads to slightly larger posterior credible intervals,  indicative of a less restrictive functional form. 

This parametrization is represented by a set of $N+1$ points $\theta = (c_{s,i}^2,\varepsilon_{i})$.  The squared speed of sound $c_{s}^2(\varepsilon, \theta)$ is modelled as a linear interpolation between these points,  i.e. for $\varepsilon \in [\varepsilon_{i}, \varepsilon_{i+1}]$ with $ i = 0, 1,\dots, N$ one has: 
\begin{eqnarray}
	c_{s}^2(\varepsilon, \theta) = \frac{(\varepsilon_{i+1} - \varepsilon) c_{s,i}^2 + (\varepsilon - \varepsilon_{i})c_{s,i+1}^2}{\varepsilon_{i+1} - \varepsilon_{i}}~.
	\label{eq:parametrization}
\end{eqnarray}
At very low densities, $n \leq 0.5 \, n_0$,  the speed of sound is matched to the well established Baym-Pethick-Sutherland (BPS) neutron star crust EoS \cite{Baym1971}.  This fixes the parameters $(c_{s,0}^2, \varepsilon_{0}) = (c_{s,\text{crust}}^2, \varepsilon_\text{crust})$.  Beyond the last point, $\varepsilon > \varepsilon_N$,  the speed of sound becomes constant, $c_{s}^2(\varepsilon, \theta) = c_{s,N}^2$.  Here we choose $N = 5$,  corresponding to a total of seven segments, which turns out to be more than enough for a representation of $c_s^2$.  In fact it was found in Ref.\,\,\cite{Annala2023} that four segments are already sufficient to describe the current astrophysical data and lead to results comparable to those of a non-parametric Gaussian process.  Note that in contrast to our previous work in Ref.\,\,\cite{Brandes2023},  the last point ($i=N$) is no longer fixed to reproduce the asymptotic conformal limit, such that the number of free parameters of the parametrization is increased to ten.  The pQCD constraint at very high densities,  far beyond those realized in the core of even the heaviest neutron stars,  will nonetheless be systematically implemented. \\ \\

A central focus of the present work is the possible occurrence of a first-order phase transition in neutron star matter.  In the EoS this corresponds to a jump in energy density,  i.e.  the appearance of two successive  discontinuities in the speed of sound.  For instance,  in a Maxwell construction a phase coexistence region of constant pressure emerges along a density interval $\Delta n$.  At the lower end of this interval the sound velocity drops to zero while at the upper end it jumps back to a finite value.  This and similar kinds of scenarios are represented in the parametrization of Eq.\,(\ref{eq:parametrization}) when one of the interpolation points reaches a small sound speed, $c_{s,i}^2 \sim 0$,  while the two adjacent points remain at finite values.  This condition as such is not sufficient to identify a first-order phase transition.  However,  in combination with a detailed quantitative inspection of $\Delta n/n$ as a measure for the extension of a phase coexistence region that can possibly develop within the posterior credible bands,  it serves to set constraints on the appearance of a strong first-order transition.  We refer to a first-order transition as `strong' if $\Delta n/n > 1$ (where $n$ is the density at which the coexistence interval starts).  In contrast a `weakly' first-order transition has $\Delta n/n$ small compared to unity.


\subsection{Bayesian inference}

For given data $\mathcal{D}$ and model $\mathcal{M}$,  the posterior probability distribution for the parameters $\theta$ is computed using Bayes' theorem:
\begin{align}
	\text{Pr} (\theta|\mathcal{D}, \mathcal{M}) \propto \text{Pr} (\mathcal{D}|\theta, \mathcal{M}) \, \text{Pr} (\theta| \mathcal{M}) ~,
	\label{eq:BayesTheorem}
\end{align} 
The prior $\text{Pr} (\theta| \mathcal{M})$ is fixed by choosing wide ranges for the parameters introduced in the previous section.  For given parameters $\theta$, the likelihood $\text{Pr} (\mathcal{D}|\theta, \mathcal{M})$ has to be computed based on the available astrophysical data. Via numerically solving the coupled system of TOV equations and the equations for the tidal deformability, a set of parameters $\theta$ is deterministically linked to $M$, $R$ and $\Lambda$, such that we can write
\begin{eqnarray}
	\text{Pr} ( \mathcal{D}|\theta, \mathcal{M}) = \text{Pr} ( \mathcal{D}|M, R, \Lambda, \mathcal{M})~.
\end{eqnarray} 
In the following we assume that the posterior distributions derived from the analyses of neutron star data can be utilized as likelihoods in the inference process:
\begin{eqnarray}
	\text{Pr} ( \mathcal{D}|M, R, \Lambda, \mathcal{M}) \propto \text{Pr} ( M, R, \Lambda |\mathcal{D}, \mathcal{M})~,
\end{eqnarray}
which is valid if the priors in $(M, R, \Lambda)$ used in the analyses of the observational data are sufficiently flat.  This is indeed the case for the observables analysed in this work \cite{Raaijmakers2021}. 

After determining the posterior probability distribution for the parameters $\theta$ according to Eq.\,(\ref{eq:BayesTheorem}),  marginal credible bands can be computed.  This is done by discretizing the pertinent variable on a grid,  e.g.  for $c_s^2(\varepsilon)$ the energy density is discretized to $\{\varepsilon_j\}$.  At each grid point $\varepsilon_j$ credible intervals are computed based on $\{c_s^2(\varepsilon_j, \theta)\}$. These are connected to obtain the marginal posterior credible bands.  Note that each EoS characterized by parameters $\theta$ is only used up to the maximum central energy density $\varepsilon_{c,\max}$,  the one corresponding to the maximum TOV mass $M_{\max}$ supported by that EoS. The latter statement applies unless the EoS leads to more than one stable branch in the mass-radius relation.

We use Bayes factors $\mathcal{B}_{H_0}^{H_1}$ to quantify the evidence for some given hypothesis $H_1$ over a counter hypothesis $H_0$.  A Bayes factor can be computed as the ratio of marginal likelihoods for the two given hypotheses.  Of particular interest is the detailed Bayes factor analysis of the evidence for or against a
rapid variation of energy density with pressure, corresponding to a
low averaged sound speed in neutron star cores.
In order to arrive at statistically well-defined conclusions,  we compare the resulting Bayes factors to a commonly used classification scheme from Refs.\,\,\cite{Jeffreys1961,Lee2016}.

\subsection{Priors}
\label{sec:BayesianInference}
\subsubsection{Sound speed prior}

To achieve maximally general results,  the ten parameters for sound speeds and energy densities, $\theta = (c_{s,i}^2,\varepsilon_{i})$,  are varied over extremely broad ranges. The energy densities are sampled logarithmically from $\varepsilon_i \in [\varepsilon_{\text{crust}},  4 \, \text{GeV} \, \text{fm}^{-3}]$.
The speed-of-sound values are collected from logarithmic intervals $c_{s,i}^2 \in [0,1]$ spanning the whole permitted range.  Thus, they are causal and thermodynamically stable by construction.  
In contrast to Ref.\,\,\cite{Brandes2023},  we do not reject parameter sets that lead to disconnected branches, hence our search includes the possibility of twin-star scenarios and even cases with more than two disconnected branches in the mass-radius relation. 

\begin{figure}[tbp]
	\begin{center}
		\includegraphics[height=55mm,angle=-00]{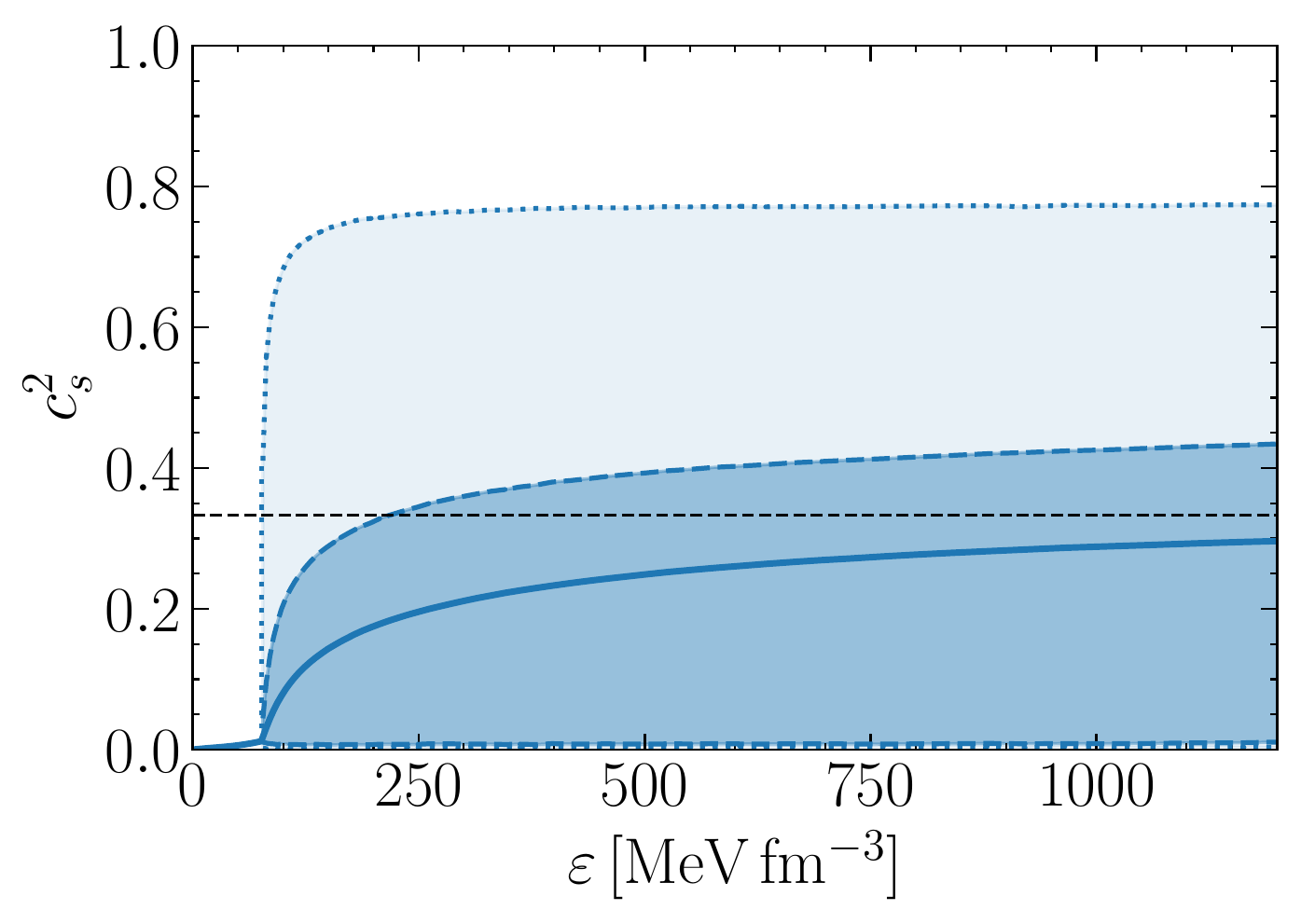}
		\caption{Marginal prior probability distributions at the 95\% and 68\% levels of the squared speed of sound $c_s^2$ as a function of energy density $\varepsilon$.  At each $\varepsilon$,  there exist 95\% and 68\% prior credible intervals for $c_s^2(\varepsilon)$. These intervals are connected to obtain the prior credible bands. The solid line represents the median of the prior distribution.  The dashed black line indicates the value $c_s^2 = 1/3$ of the conformal bound.}
		\label{fig:Prior}
	\end{center}
\end{figure}

The prior credible bands for the speed of sound as a function of energy density are displayed in Fig.\,\,\ref{fig:Prior}. With a 95\% credible band that spans a wide range of sound speeds,  it is apparent that the prior probability distribution is quite unrestrictive.  Due to the logarithmic sampling of the $c_{s,i}^2$,  the 68\% prior credible band resides at smaller sound speeds. Accordingly, 15\% of the EoSs in the prior distribution display a possible strong phase transition, represented in our
parametrisation by a minimum speed of sound $c_{s,\min}^2 \leq 0.1$.  In contrast to the posterior credible bands,  for the computation of the prior credible bands,  each EoS is employed unrestrictedly up to arbitrary $\varepsilon > \varepsilon_{c,\max}$.

\subsubsection{Mass prior}

Each EoS characterized by a parameter set $\theta$ supports neutron star masses between some minimum mass $M_\text{min}$ and a respective maximum mass $M_{\max}(\theta)$.  We follow Ref.\,\,\cite{Landry2020} and assume a flat prior distribution between $M_\text{min}$ and $M_{\max}(\theta)$:
\begin{align}
	\text{Pr}(M(\theta)) = \begin{cases}
		\frac{1}{M_{\max}(\theta) - M_\text{min}} &\text{ if } M\in[M_\text{min}, M_{\max}(\theta)] \\
		0 &\text{ else}
	\end{cases}   
	\label{eq:MassPrior}
\end{align}
with a minimum mass of $M_\text{min} = 0.5\, M_\odot$. This introduces an Occam factor penalising EoSs that involve extreme masses beyond those supported by the astrophysical data.  

The employment of a uniform mass prior differs from our previous work \cite{Brandes2023} where a central pressure prior was used instead,  referring to \cite{Miller2019a,Greif2019}.  In Ref.\,\,\cite{Gorda2023} the authors found just a marginal distinction between a flat mass prior and a central pressures prior.  However,  the uniform mass prior has the advantage that it permits a more direct comparison of our results with other recent works \cite{Biswas2021,Biswas2022,Gorda2023,Legred2021,Takatsy2023,Han2023,Lim2022,Essick2020,Landry2020,Essick2023}.  Furthermore,  for future developments it offers the possibility of easily incorporating the mass population of neutron stars.  When the number of available data increases,  failing to account for the correct population model may cause a bias in the resulting posterior distributions \cite{Mandel2019}. 

\section{Likelihoods,  data and constraints}
\label{sec:likelihoods}

In order to compute the posterior probability distribution with Eq.\,(\ref{eq:BayesTheorem}) it is necessary to compute likelihoods for the different types of data.  In this section a brief account is given of the sets of empirical data used in the inference procedure. This also includes the recently explored heavy black widow pulsar PSR J0952-0607 and the 
extremely light supernova remnant HESS J1731-347.  Both of these objects were not yet available for assessment in our previous publication \cite{Brandes2023}.  Finally,  the low-density constraint from ChEFT and the high-density matching to pQCD will be shortly reviewed and re-examined.

\subsection{Shapiro time delay measurements}

Shapiro time delay refers to the shift of the measured pulsar frequency signal in the presence of a white dwarf companion.  From this delay the mass of the pulsar can be extracted with high precision. Using this method several neutron stars with masses around twice the solar mass were established, namely PSR J1614–2230 \cite{Demorest2010,Fonseca2016,Arzoumanian2018}, PSR J0348+0432 \cite{Antoniadis2013} and PSR J0740+6620 \cite{Cromartie2020,Fonseca2021}. The respective extracted masses are listed in Tab.\,\,\ref{tab:DataSet}. To compute the likelihoods for this sort of data, we assume that the resulting mass probability distributions are Gaussians, $\mathcal{N}\left(M, \langle M \rangle, \sigma_M\right) = (1/\sqrt{2\pi \sigma_M^2}) \,\text{exp}[-1/2 \, (M - \langle M \rangle)^2/\sigma_M^2]$ with mean values $\langle M \rangle$ and standard deviations $\sigma_M$. Then the likelihood is equivalent to the integral over the Gaussian probability distribution, weighted with the mass prior in Eq.\,(\ref{eq:MassPrior}):
\begin{align}
	\text{Pr} \big(M(\theta)&\big|\mathcal{D}_{\text{Shapiro}}, \mathcal{M}\big) \nonumber \\
	& = \int_{M_{\min}}^{M_{\max}(\theta)} \text{d}M \,~ \mathcal{N}\left(M, \langle M \rangle, \sigma_M\right) \, \text{Pr}(M(\theta)) \nonumber \\ 
	& \approx \frac{1}{2} \left[1  + \text{erf}\left(\frac{M_{\max}(\theta) - \langle M \rangle}{\sqrt{2}\sigma_M}\right)\right] \, \text{Pr}(M(\theta))~.
	\label{eq:LikelihoodShapiro}
\end{align} 

\begin{table}[tbp]
	\centering
	\begin{tabularx}{\linewidth}{|l|Xl|}
		\hline \hline
		\multicolumn{3}{|l|}{Data and constraints}  \\ \hline
		& PSR J1614–2230 & $M = 1.908 \pm 0.016 \, M_\odot$ \cite{Arzoumanian2018} \\
		& PSR J0348+0432 & $M = 2.01 \pm 0.04 \, M_\odot$ \quad \cite{Antoniadis2013} \\
		& PSR J0030+0451 & $M=1.34^{+0.15}_{-0.16} \, M_\odot$  \\
		&& $R =12.71^{+1.14}_{-1.19}\,\text{km}$ \quad \cite{Riley2019} \\
		& PSR J0740+6620 & $ M = 2.072^{+0.067}_{-0.066} \, M_\odot$  \\
		Previous && $R = 12.39^{+1.30}_{-0.98}\,\text{km}$ \quad \cite{Riley2021} \\
		& GW170817 & $\tilde{\Lambda}=320^{+420}_{-230}$ \quad \cite{Abbott2019} \\
		& GW190425 & $\tilde{\Lambda} \leq 600$ \quad \cite{Abbott2020} \\ 
		& ChEFT & \cite{Drischler2021,Drischler2022} \\
		& pQCD & \cite{Gorda2021,Komoltsev2022,Gorda2023} \\ [.8ex]
		\hline
		BW & \multicolumn{1}{l}{PSR J0952-0607} & $M = 2.35 \pm 0.17 \, M_\odot$  \cite{Romani2022} \\ [.8ex] \hline 
		HESS & \multicolumn{1}{l}{HESS J1731-347} & $M =0.77^{+0.20}_{-0.17} \, M_\odot$ \\
		&& $R = 10.4^{+0.86}_{-0.78}\,\text{km}$ \quad \cite{Doroshenko2022} \\ [.8ex]
		\hline \hline  
	\end{tabularx}
	\caption{Data and constraints used in this Bayesian inference analysis.  We examine the impact of the new black widow (BW) and supernova remnant (HESS) data separately from the previously available data (Previous).  All listed results are at the 68\% level except for the $\tilde{\Lambda}$ results based on the gravitational wave measurements which are at the 90\% level.}
	\label{tab:DataSet}
\end{table}

\subsection{X-ray measurements: pulse profile modelling}

Through the analysis of x-ray emission from hot spots on the surfaces of rapidly rotating pulsars,  posterior distributions for the mass and radius of selected neutron stars have been derived.
Two x-ray spectra of neutron stars were measured so far by the NICER telescope on board the ISS.  Here we use the results from the analysis of the NICER data by Riley \textit{et al.} \cite{Riley2019,Riley2021} listed in Tab.\,\,\ref{tab:DataSet}.  The complementary NICER data analysis by Miller \textit{et al.} found slightly different results \cite{Miller2019,Miller2021} but they are consistent within uncertainties.  Based on our previous studies we expect the choice between these two analyses to have only little influence on the conclusions \cite{Brandes2023}. The posterior probabilities for the mass and radius data are available as samples and we can approximate the underlying distributions using Kernel Density Estimation (KDE).  For a given EoS characterised by parameters $\theta$, we solve the TOV equations to obtain $R(M,\theta)$. The likelihood is then computed as the mass integral over the KDE evaluated at the radius given by the mass-radius relation: 
\begin{align}
	\text{Pr}\big( (M,R)&(\theta)  \big| \mathcal{D}_{\text{NICER}}, \mathcal{M}\big) \nonumber\\  
	& = \int_{M_{\min}}^{M_{\max}(\theta)} \text{d}M \,~ \text{KDE}\big(M,R(M,\theta)\big) \text{Pr}(M(\theta)) ~. 
	\label{eq:LikelihoodNICER}
\end{align}
This is again weighted with the mass prior in Eq.\,(\ref{eq:MassPrior}).  For PSR J0740+6620,  both Shapiro time delay and NICER measurements are available.  In this case we do not include the Shapiro mass data in the total likelihood to avoid double counting.

\subsection{Binary neutron star mergers}

The merging of two neutron stars in a binary produces gravitational waves (GW). The signals detected by the LIGO Scientific and Virgo Collaborations can be compared to gravitational waveform models.  From these observations a posterior for the masses $(M_1, M_2)$ and tidal deformabilities $(\Lambda_1, \Lambda_2)$ of the two neutron stars can be inferred. So far,  two possible binary neutron star merger events,  GW170817 \cite{Abbott2019} and GW190425 \cite{Abbott2020},  have been detected. The resulting mass-weighted combinations of tidal deformabilities, $\tilde{\Lambda}$,  based on the analysis of both GW events are listed in Tab.\,\,\ref{tab:DataSet}. We can again approximate the underlying probability distribution of the available posterior samples via Kernel Density Estimation. For a given set of parameters $\theta$,  solving the TOV equations in combination with the equations for the tidal deformability yields the relationship $\Lambda(M,\theta)$.  Masses larger than the maximum mass of a given EoS, $M > M_\text{max}(\theta)$,  correspond to a black hole,  in which case $\Lambda(M,\theta)$ is set to zero. The likelihood is given by the integral over both neutron star masses,  with the tidal deformabilities given by $\Lambda(M,\theta)$:  
\begin{align}
	\text{Pr}&\big( (M,\Lambda)(\theta)  \big| \mathcal{D}_{\text{GW}}, \mathcal{M}\big) = \nonumber\\  &\int \text{d}M_1 \int \text{d}M_2  \,~ \text{KDE}\big(M_1,M_2,\Lambda(M_1,\theta), \Lambda(M_2,\theta)\big) ~. 
	\label{eq:LikelihoodGW}
\end{align}
Following Ref.\,\,\cite{Landry2020} we do not use the mass prior here,  as we cannot start a priori from assuming the events to be neutron star-neutron star mergers,  but should also allow for the principal possibility of neutron star-black hole binaries.

The chirp mass, $M_\text{chirp} = (M_1M_2)^{3/5}(M_1 + M_2)^{-1/5}$, of the GW170817 event has a very small uncertainty, $M_\text{chirp} = 1.186 \pm 0.001\,M_\odot$,  hence some analyses assumed it to be fixed.  For a given $M_1$ this allows to determine $M_2$ and effectively removes one of the integrations.  However, this does not work for GW190425 where the chirp mass has a larger uncertainty, $M_\text{chirp} = 1.44 \pm 0.02\,M_\odot$.  Moreover,  determining $M_2$ via $M_\text{chirp}$ neglects the $M_1$ dependence of the chirp mass. We perform the double integration in Eq.\,(\ref{eq:LikelihoodGW}) over both $M_1$ and $M_2$ without resorting to a fixed chirp mass.

\subsection{Black widow pulsar PSR J0952-0607}

The heaviest neutron star observed so far was recently reported in Ref.\,\,\cite{Romani2022}.  The pulsar PSR J0952-0607 is estimated to have a total mass $M = 2.35 \pm 0.17 \, M_\odot$,  significantly larger than previously observed masses based on Shapiro time delays,  but with a relatively large uncertainty. This object is a so-called black widow pulsar,  meaning that the neutron star accreted much of its mass from a lighter companion.  With a rotational frequency $\nu = 709\,$Hz,  PSR J0952-0607 is also among the fastest-spinning pulsars. Therefore rotation corrections have to be considered as they can effectively increase the maximum mass that a given EoS can support.  We use the radius-dependent rotation adjustment provided in Ref.\,\,\cite{Konstantinou2022},  where the authors find approximately universal relations between stationary and rotating masses and radii. They fit an empirical formula,  independent of the EoS,  to derive the mass and radius corrections induced by a rotation with frequency $\nu$. For the resulting radius-dependent rotation correction of the PSR J0952-0607 mass see Appendix \ref{sec:RotationAdjustment}. In order to compute the black-widow likelihood for a given set of parameters $\theta$, we first determine the TOV maximum non-rotating mass $M_\text{max}(\theta)$ as well as the corresponding radius $R(M_\text{max}, \theta)$.   With the formulas derived in Ref.\,\,\cite{Konstantinou2022} we can then compute the rotation-adjusted maximum mass $M_\text{rot,max}(\theta)$.  Assuming that the mass distribution of PSR J0952-0607 is Gaussian,  its likelihood can be computed in a way similar to the Shapiro time delay likelihood in Eq.\,(\ref{eq:LikelihoodShapiro}),  with the non-rotating maximum mass replaced by $M_\text{rot,max}(\theta)$. 

In addition to PSR J0952-0607 there exits another similarly massive object, PSR J2215+5135, a redback pulsar with a reported mass of $M = 2.27_{-0.15}^{+0.17}\,M_\odot$ \cite{Linares2018}. Like other black-widow pulsars, this object required a more complex heating model which introduces a larger systematic uncertainty \cite{Linares2018,Romani2022,Kandel2022}. It is therefore not included in the present analysis.  

\subsection{Supernova remnant HESS J1731-347}
\label{sec:LikelihoodHESS}

The observation of a neutron star with an unusually small mass, a central compact object within the supernova remnant HESS J1731-347, was recently reported \cite{Doroshenko2022}. Such central compact objects have weak magnetic fields and nearly constant thermal x-ray emission.  In their analysis, the authors find a mass of only $M=0.77^{+0.20}_{-0.17} \, M_\odot$ as well as a small radius,  $R=10.4^{+0.86}_{-0.78}\,\text{km}$ \cite{Doroshenko2022}. This low mass is remarkable because it is not clear how neutron stars with masses lower than around $1.17\, M_\odot$ form based on known neutron star evolution mechanisms that involve supernovae \cite{Suwa2018}.  The previously known lightest neutron star is compatible with this low mass constraint \cite{Tauris2019}. This in combination with the object's small radius lead the authors to speculate that HESS J1731-347 might be a possible strange star.  Because of the absence of pulsations in the measured spectra, the authors assumed a uniform surface temperature as well as a carbon atmosphere.  In addition, they assume that the stellar magnetic field has no surface effect.  However,  other authors have suggested a non-uniform emission for similar kinds of central compact objects \cite{Alford2023}.  In that case larger masses and radii might be possible for HESS J1731-347.  Despite such obvious model dependence in the interpretation of the data,  it is instructive to add HESS J1731-347 as a separate item in our inference analysis,  just in order to explore what its impact would be on the overall picture.  The mass-radius posterior from Ref.\,\,\cite{Doroshenko2022} is again available as samples,  so we can compute the likelihood in a way similar to the NICER analyses in Eq.\,(\ref{eq:LikelihoodNICER}).

\subsection{Low-density constraint:\\ chiral effective field theory}

Chiral effective field theory (ChEFT) is the method of choice to calculate nuclear many-body systems  at low baryon densities. The low energy constants of this effective field theory are fitted to a large number of empirical nucleon-nucleon and pion-nucleon interaction data. The approach is then extended to many-body systems at finite densities.  In recent works a novel ansatz has been introduced based on a Gaussian Process to derive combined uncertainties from many-body approximations and from missing higher order terms beyond next-to-next-to-next-to-leading chiral order (N3LO) \cite{Melendez2019,Drischler2020,Drischler2020a}.  In this way posterior credible bands were derived for the squared speed of sound,  $c_{s}^2(n)$,  as a function of density up to two times nuclear saturation density,  $n = 2\,n_0$ \cite{Drischler2021,Drischler2022}.  These credible bands also agree with the results found independently by another group \cite{Keller2023}. 

In many other recent Bayesian studies,  the ChEFT constraints were implemented as a prior.  However,  the ChEFT framework with its low-energy constants and uncertainty measures should be considered as representing a large variety of empirical nuclear physics data,  in the same general category as the astrophysical data. There is,  in principle,  no reason to trust the uncertainty estimates of ChEFT more than those of the astrophysical data.  We therefore follow our previous work \cite{Brandes2023} and employ the ChEFT information as a likelihood instead of a prior.  The likelihood treatment permits a balancing between the constraints from nuclear physics and astrophysics and allows for a rigorous and statistically consistent Bayes factor analysis.

With the Gaussian Process used in Refs.\,\,\cite{Drischler2021,Drischler2022},  the speed of sound is normally distributed at each density $n$ with mean value $\langle c_s^2(n)\rangle$ and standard deviation $\sigma(n)$. We employ the ChEFT results at discrete densities $n_i$ starting from the BPS crust and extending up to a maximum density for the applicability of the effective field theory,  $n_i \in [0.5\,n_0, n_\text{ChEFT}]$.  The ChEFT likelihood is then computed via Bayesian linear regression,  that is the total likelihood is given by the product of the Gaussian likelihoods at each $n_i$:
\begin{align}
	\text{Pr}&\big( \mathcal{D}_{\text{ChEFT}} \big| c_s^2(n,\theta),  \mathcal{M}\big) \nonumber\\ 
	& \propto \prod_i \exp\left[-\frac{1}{2} \,~ \left(\frac{\langle c_s^2(n_i)\rangle - c_s^2(n_i, \theta)}{\sigma(n_i)}\right)^{2} \right]  \nonumber \\
	& = \exp\left[-\frac{1}{2} \sum_i \,~ \left(\frac{\langle c_s^2(n_i)\rangle - c_s^2(n_i, \theta)}{\sigma(n_i)}\right)^{2} \right] ~, \nonumber \\
\end{align}
where we assign each density the same prior weight.  The speed-of-sound constraint of Refs.\,\,\cite{Drischler2021,Drischler2022} is continuous in $n$. Therefore we choose to replace the sum by an integral:
\begin{align}
	\text{Pr}&\big( \mathcal{D}_{\text{ChEFT}} \big| c_s^2(n,\theta),  \mathcal{M}\big) \nonumber\\ 
	& \propto \exp\left[-\frac{1}{2}\int_{0.5\,n_0}^{n_\text{ChEFT}} \text{d}n  \,~ \left(\frac{\langle c_s^2(n)\rangle - c_s^2(n, \theta)}{\sigma(n)}\right)^{2} \right] ~.
	\label{eq:LikelihoodChEFT}
\end{align}
We have checked that this likelihood leads to posterior credible bands for the sound velocity that are very similar to those in Refs.\,\,\cite{Drischler2021,Drischler2022}.  Following the results of the analysis in Ref.\,\,\cite{Essick2020}, we choose a conservative maximum density for the applicability of ChEFT as $n_\text{ChEFT} = 1.3\,n_0$.  At higher densities the N2LO and N3LO results become increasingly different,  hinting towards possible convergence issues. We will examine the impact of this choice in the later analysis.

\subsection{High-density matching: perturbative QCD}

At asymptotically high densities, $n \gtrsim 40\,n_0$, far beyond those encountered in neutron stars,  the QCD coupling turns sufficiently weak such that perturbative calculations become feasible.  As pointed out recently \cite{Komoltsev2022,Gorda2023},  even though pQCD is applicable only at extremely high densities,  demanding that any valid EoS should be connected to these asymptotic pQCD results can lead to constraints at much lower densities. 

The partial N3LO pQCD results of Ref.\,\,\cite{Gorda2021} are taken to be valid at the chemical potential $\mu_\text{pQCD} = 2.6\,$GeV, with corresponding density $n_\text{pQCD}$ and pressure $P_\text{pQCD}$.  For any point of a given equation of state,  $n_\text{NS}$, $P_\text{NS}$ and $\mu_\text{NS}$,  it must be possible to connect to the asymptotic pQCD constraint via a causal and thermodynamically stable interpolation.  Because the squared speed of sound (derived in the grand canonical approach from $P(\mu)$) is causally limited,
\begin{align}
	c_s^2 = \left(\frac{\mu}{n}\frac{\partial n}{\partial \mu }\right)^{-1} \leq 1 ~,
\end{align}
a minimal slope of the function $n(\mu)$ is determined for any specific EoS:
\begin{align}
	\frac{\partial n}{\partial \mu } \geq \frac{n}{\mu} ~.
\end{align}
Demanding that it should be possible to connect a point in the neutron star range,  $(\mu_\text{NS}, n_\text{NS}, P_\text{NS})(\theta)$,  to $\mu_\text{pQCD}$, $n_\text{pQCD}$ and $P_\text{pQCD}$ leads to the integral constraint
\begin{align}
	\int_{\mu_\text{NS}}^{\mu_\text{pQCD}} d\mu \,~ n(\mu) = P_\text{pQCD} - P_\text{NS} = \Delta P ~.
\end{align}
It was shown in Ref.\,\cite{Komoltsev2022} that the requirements of causality and thermodynamic stability imply the following minimum and maximum values for $\Delta P$:
\begin{align}
	\Delta P_{\min} &= \frac{\mu_\text{pQCD}^2 - \mu_\text{NS}^2}{2\mu_\text{NS}}\,n_\text{NS} \\
	\Delta P_{\max} &= \frac{\mu_\text{pQCD}^2 - \mu_\text{NS}^2}{2\mu_\text{pQCD}}\,n_\text{pQCD} ~.
\end{align}
Accordingly,  the pQCD likelihood is equal to one if the difference $\Delta P$ is within these two values and zero otherwise:
\begin{align}
	\text{Pr}\big( \mathcal{D}_{\text{pQCD}}  & \big| \Delta P(\theta), \mathcal{M}\big) \nonumber \\ 
	& = \begin{cases}
		1 & \text{if } \Delta P(\theta) \in [\Delta P_{\min}(\theta), \Delta P_{\max}(\theta)] \\
		0 & \text{else} 
	\end{cases} ~. 
	\label{eq:LikelihoodpQCD}
\end{align}
The pQCD results still depend on a renormalisation scale $X$.  We follow Refs.\,\,\cite{Komoltsev2022,Gorda2023} and take the logarithmic average over $X \in [1/2, 2]$.  Each EoS is constrained by neutron star data only up to the respective maximum central chemical potential $\mu_{c,\max}$, density $n_{c,\max}$ and pressure $P_{c,\max}$.  As in Refs.\,\cite{Somasundaram2023a,Takatsy2023} we verify at that point,  $(\mu_\text{NS}, n_\text{NS}, P_\text{NS}) = (\mu_{c,\max}, n_{c,\max}, P_{c,\max})$,  whether a causal and thermodynamic interpolation to the asymptotic pQCD constraint exists.  Note that the authors of Refs.\,\cite{Komoltsev2022,Gorda2023} chose $n_\text{NS} = 10\,n_0$ instead,  together with the corresponding chemical potential and pressure,  well above the central densities reached in neutron stars.  We will analyse the impact of this choice in the later analysis.

The set consisting of Shapiro time delay data,  NICER measurements and the information from gravitational wave events,  together with ChEFT and pQCD constraints,  is denoted 'Previous' in the following.  This serves to study the impact of the two new observations,  i.e. the black widow (BW) mass and the supernova remnant (HESS) mass-radius data.  This nomenclature is also displayed in Tab.\,\,\ref{tab:DataSet}, where we summarize all data used in the present Bayesian analysis.

\section{RESULTS}
\label{sec:Results}

This section starts with a presentation of the posterior results for neutron star properties (masses, radii, central densities) together with the posterior credible bands for the sound speed and the equation-of-state.  This is done first for the "previous" data base and then with further incorporation of the "new" black widow PSR J0952-0607 mass information.  Thereafter we focus on detailed Bayes factor investigations, with special emphasis on low sound speeds in combination with extended phase coexistence regions in neutron star matter.  Further issues are the assessment of twin star scenarios and the discussion of the trace anomaly measure with its related quest for the approach to conformal matter.  The impacts of the low-density (ChEFT) and high-density (pQCD) constraints on the inference procedure will be carefully examined.  The section ends with comments on possible effects on the overall systematics induced by the ultralight HESS J1731-347 supernova remnant. 

\subsection{Neutron star properties}

By combining the likelihoods for the astrophysical data with the constraints from ChEFT and pQCD introduced in the previous section, we compute the total likelihood and, using Eq.\,(\ref{eq:BayesTheorem}), the posterior probability distribution for the parameters $\theta$. \newpage

\begin{figure}[tbp]
	\begin{center}
		\includegraphics[height=55mm,angle=-00]{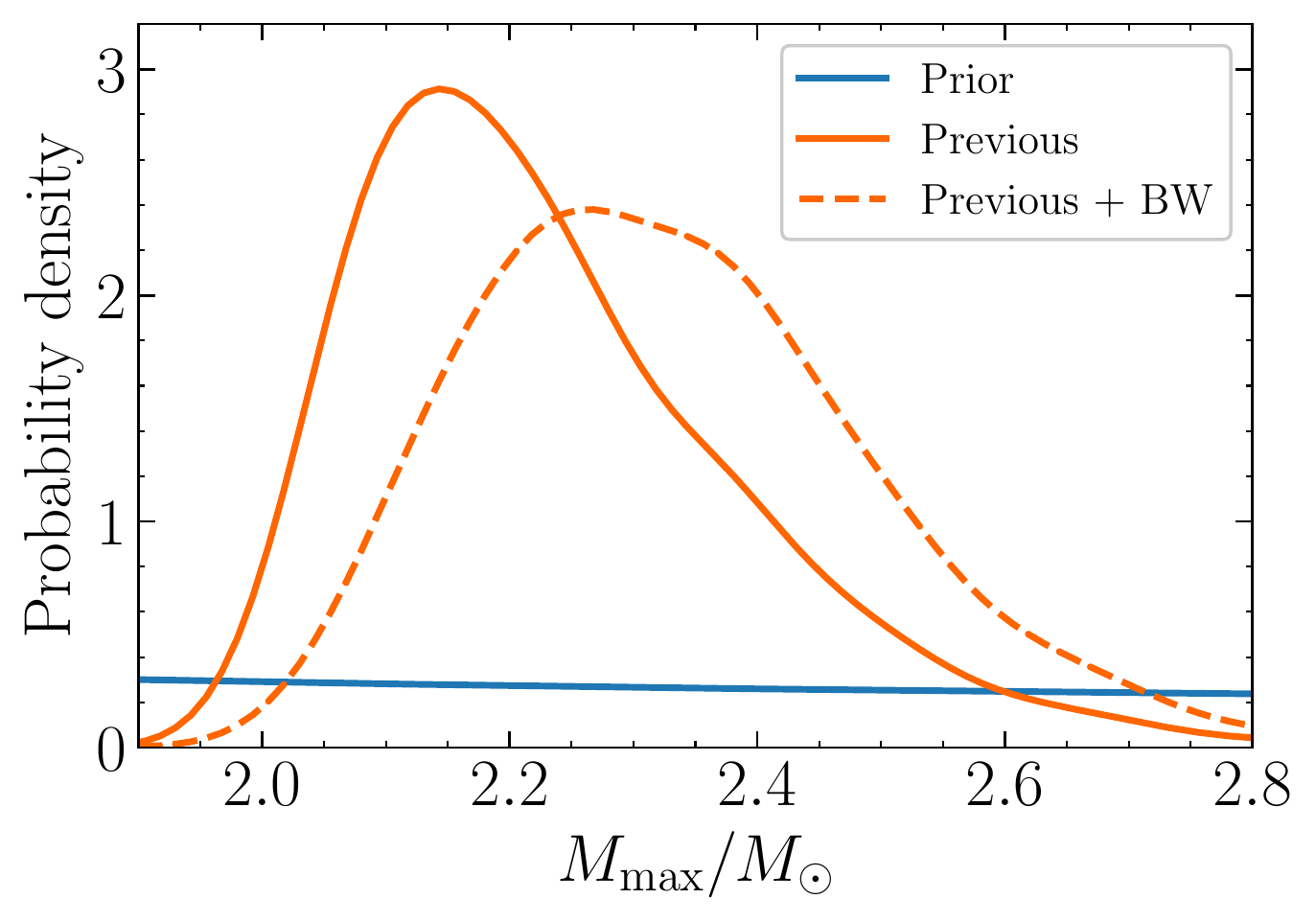} \\
		\caption{Marginal posterior probability distributions of the maximum neutron star mass $M_\text{max}$. The nomenclature 'Previous' refers to the Shapiro time delay, NICER, ChEFT, pQCD and gravitational wave data listed in Tab.\,\,\ref{tab:DataSet}.  The shift induced by adding the (non-rotating) mass information from the black widow pulsar PSR J0952-0607 (BW) is also shown. The prior distribution is nearly uniform over a wide mass range.}
		\label{fig:MaximumMass}
	\end{center}
\end{figure}

Marginalising this posterior with respect to the maximum mass $M_{\max}$, we find median and 68\% credible interval values of $M_{\max} = 2.20_{-0.16}^{+0.10}\,M_\odot$ using the 'Previous' data set.  (If not stated otherwise, from here on we always report medians and 68\% credible intervals in the text.) The corresponding probability distribution is displayed in Fig.\,\,\ref{fig:MaximumMass}. The upper limit of the 68\% credible interval extends to higher masses compared to the result of our previous analysis in Ref.\,\,\cite{Brandes2023} because of the different likelihood designs.  Recent studies using comparable likelihoods as the present one find similarly large maximum masses \,\,\cite{Legred2021,Gorda2023}.  The maximum central neutron star density turns out to be $n_{c,\max} = 6.0_{-0.8}^{+0.7}\,n_0$.  Changing the likelihood to the version used in our previous work which includes information from the ChEFT constraint up to $n \sim 2.0\,n_0$,  we find a larger central density of $n_{c,\max} = 6.4 \pm 0.6\,n_0$, in agreement with the value reported in Ref.\,\,\cite{Brandes2023}.  

Including the new mass information of the black widow pulsar PSR J0952-0607 in the data base,  the maximum non-rotating neutron star mass increases significantly to $M_{\max} = 2.31_{-0.17}^{+0.14}\,M_\odot$.  The median and the 68\% credible interval extend to masses lower than that of the PSR J0952-0607 mass,  $ 2.35\,M_\odot$.  This is due to the large mass uncertainty and the rotation correction that we have applied in the analysis.  To support such higher masses the sound speed needs to increase more rapidly,  implying a stiffer EoS which in turn leads to smaller central densities in the neutron star core.  Accordingly,  the maximum central density decreases to $n_{c,\max} = 5.6 \pm 0.7\,n_0$ with inclusion of the new heavy-mass data. 

Tab.\,\,\ref{tab:NS_properties1} collects median and credible intervals of selected inferred properties of neutron stars with masses of $M = 1.4\, M_\odot$ and $2.1\, M_\odot$.  With the 'Previous' data set but including the updated ChEFT likelihood we find $n_c = 4.1_{-0.9}^{+0.8}\,n_0$ for the central density of a $2.1\,M_\odot$ neutron star,  comparable to the value for a $2.0\,M_\odot$ neutron star reported in Ref.\,\,\cite{Legred2021}. The radius of a $2.1\,M_\odot$ neutron star, $R = 11.9 \pm 0.7\,$km, agrees with the value for $R(M = 2.0\,M_\odot)$ reported in Ref.\,\,\cite{Altiparmak2022}.

The inclusion of the new information from PSR J0952-0607 has only a marginal impact on the properties of neutron stars with relatively small masses.  However,  for a $2.1\,M_\odot$ neutron star, the inclusion of the heavy mass information reduces the central density significantly to $n_c = 3.6\pm 0.7\,n_0$.  Similarly,  the central energy density $\varepsilon_c$ and central pressure $P_c$ are reduced,  while the radius $R$ is slightly increased and the tidal deformability $\Lambda$ remains almost unchanged within the 68\% credible interval. Uncertainties will eventually be further reduced once additional merger events are detected during the fourth observation run of LIGO, Virgo and KAGRA \cite{Colombo2022,Landry2020}.  In addition, four more objects are set to be measured by the NICER telescope \cite{Greif2019,Watts2019}.  Moreover,  a moment-of-inertia measurement of the neutron star PSR J0737-3039 in a double pulsar system is expected to become available in the next few years (see e.g. Ref.\,\,\cite{Landry2018} and references therein).

With the heaviest observed pulsar in mind,  Tab.\,\,\ref{tab:NS_properties2} displays inferred properties of a neutron star with mass $M = 2.3\,M_\odot$.  For such a heavy-mass object the central density is still only $n_c = 3.8_{-0.8}^{+0.7}\,n_0$,  comparable to that of a $2.1\,M_\odot$ neutron star in Tab.\,\,\ref{tab:NS_properties1}.  This result is of some significance because it indicates that the baryon densities in the cores of even the heaviest neutron stars are not expected to reach extreme values.  Under such conditions the average distance between baryons still exceeds 1$\,$fm,  more than twice the typical hard-core radius usually associated with the short-range nucleon-nucleon interaction. An interpretation of the dense matter in neutron star cores primarily in terms of nucleon degrees of freedom is therefore not excluded. Recent work by Benhar \cite{Benhar2023} comes to a similar conclusion based on the emergence of $y$-scaling and the role of short-range correlations in electron-nucleus scattering.

\setlength{\extrarowheight}{6pt}
\begin{table}[tbp]
	\centering
	\begin{tabularx}{\linewidth}{|l|l|Xl|Xl|}
		\hline \hline 
		\multicolumn{2}{|l|}{} & \multicolumn{2}{l|}{Previous} & \multicolumn{2}{l|}{Previous + BW} \\
		\multicolumn{2}{|l|}{} & 95\% & 68 \% & 95\% & 68 \% \\ \hline
		&$n_c/n_0$  & $2.8_{-0.7}^{+0.8}$ & $\pm 0.4$ & $2.6 \pm 0.7$ & $_{-0.4}^{+0.3}$ \\
		&$\varepsilon_c  \, $[MeV$\,$fm$^{-3}$] & $451_{-123}^{+133}$ & $_{-71}^{+62}$ & $423_{-116}^{+118}$ & $_{-67}^{+56}$ \\
		$1.4\, M_\odot$& $P_c \, $[MeV$\,$fm$^{-3}$] & $64_{-23}^{+30}$ & $_{-16}^{+12}$ & $60_{-20}^{+28}$ & $_{-14}^{+11}$ \\
		&$R \, $[km] & $12.2_{-1.0}^{+0.9}$ & $\pm 0.5$ & $12.3_{-1.0}^{+0.8}$ & $\pm0.5$ \\
		&$\Lambda$ & $396_{-197}^{+226}$ & $_{-127}^{+107}$  & $421_{-200}^{+236}$ & $_{-124}^{+114}$ \\ [.8ex] \hline 
		&$n_c /n_0$ & $4.1_{-1.5}^{+1.9}$ & $_{-0.9}^{+0.8}$ & $3.6_{-1.3}^{+1.6}$ & $\pm 0.7$ \\
		&$\varepsilon_c  \, $[MeV$\,$fm$^{-3}$] & $716_{-326}^{+416}$ & $_{-213}^{+162}$ & $628_{-251}^{+357}$ & $_{-146}^{+149}$ \\
		$2.1\, M_\odot$& $P_c \, $[MeV$\,$fm$^{-3}$] & $225_{-134}^{+239}$ & $_{-110}^{+62}$ & $186_{-104}^{+184}$ & $_{-80}^{+52}$ \\
		&$R \, $[km] & ${11.9 \pm 1.3}$ & $\pm 0.7$ & $12.1_{-1.2}^{+1.3}$ & $_{-0.8}^{+0.6}$ \\
		&$\Lambda$ & $21_{-15}^{+30}$ & $_{-13}^{+9}$ & $26_{-20}^{+30}$ & $_{-14}^{+10}$ \\ [.8ex]
		\hline \hline 
	\end{tabularx}
	\caption{Median, 95\% and 68\% credible intervals for selected neutron star properties given the previously available data and the new information from the black widow (BW) pulsar. These are computed from the one-dimensional posterior probability distribution marginalised over all other parameters. Listed are the central density $n_c$, central energy density $\varepsilon_c$, central pressure $P_c$, radius $R$ and tidal deformability $\Lambda$ of neutron stars with masses $M = 1.4 \, M_\odot$ and $2.1 \, M_\odot$.}
	\label{tab:NS_properties1}
\end{table}
\setlength{\extrarowheight}{4pt}

\setlength{\extrarowheight}{6pt}
\begin{table}[tbp]
	\centering
	\begin{tabularx}{\linewidth}{|l|X|lXX|}
		\hline \hline 
		\multicolumn{2}{|l|}{} & \multicolumn{3}{l|}{\hspace*{1mm} Previous + BW} \\
		\multicolumn{2}{|l|}{} && 95\% & 68 \% \\ \hline
		&$n_c /n_0$ && $3.8_{-1.3}^{+1.6}$ & $_{-0.8}^{+0.7}$ \\
		&$\varepsilon_c  \, $[MeV$\,$fm$^{-3}$] && $673_{-268}^{+363}$ & $_{-180}^{+140}$ \\
		$2.3\, M_\odot$& $P_c \, $[MeV$\,$fm$^{-3}$] && $237_{-134}^{+226}$ & $_{-104}^{+69}$ \\
		&$R \, $[km] && $12.3\pm1.2$ & $_{-0.6}^{+0.7}$ \\
		&$\Lambda$ && $14_{-10}^{+17}$ & $_{-9}^{+4}$ \\ [.8ex]
		\hline \hline 
	\end{tabularx}
	\caption{Same as Tab.\,\,\ref{tab:NS_properties1}, but median and credible intervals for a neutron star with mass $M = 2.3 \, M_\odot$ are displayed given the previously available data plus new information from the black widow (BW) pulsar PSR J0952-0607.}
	\label{tab:NS_properties2}
\end{table}
\setlength{\extrarowheight}{4pt}

\begin{figure*}[tbp]
	\begin{center}
		\includegraphics[height=55mm,angle=-00]{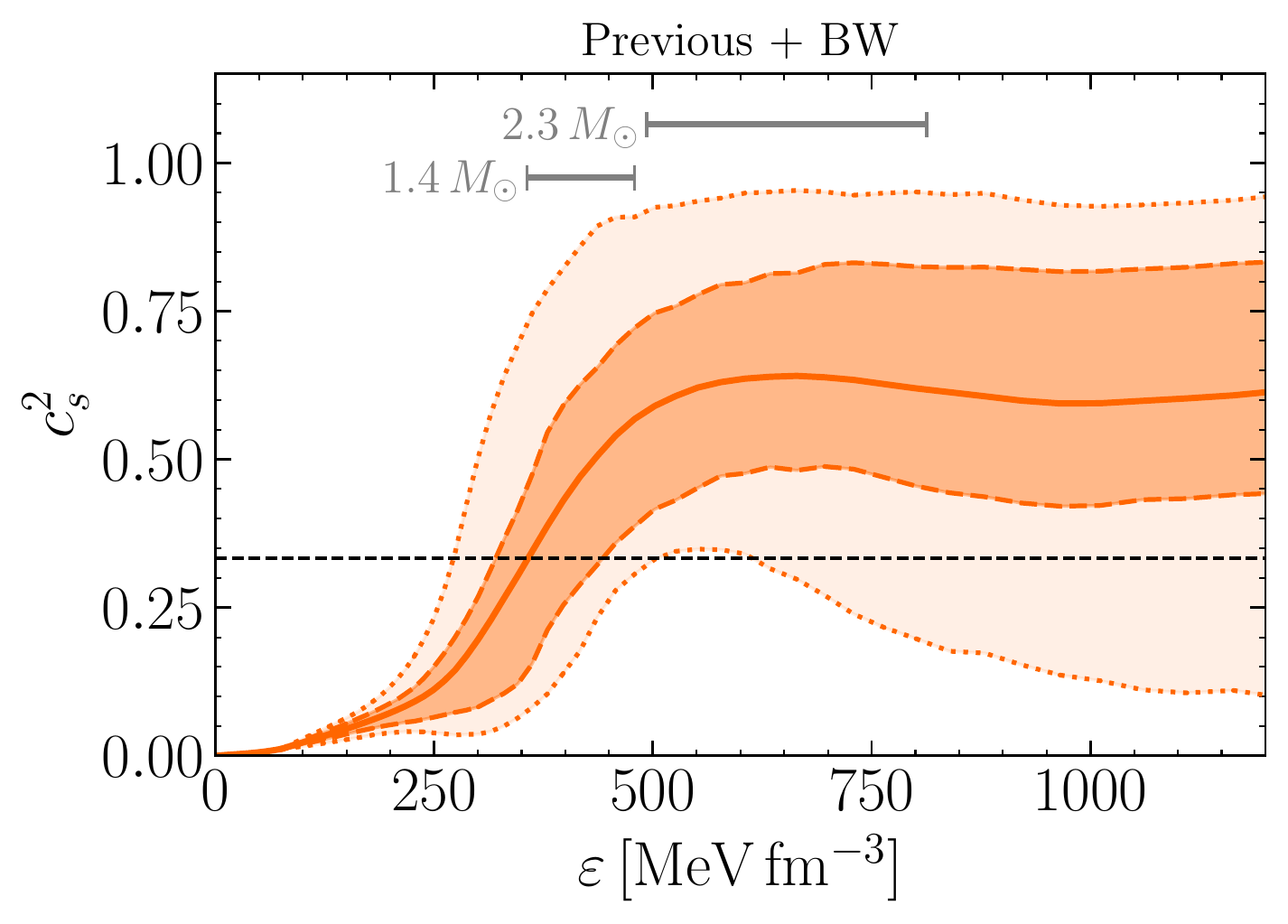} 
		\includegraphics[height=55mm,angle=-00]{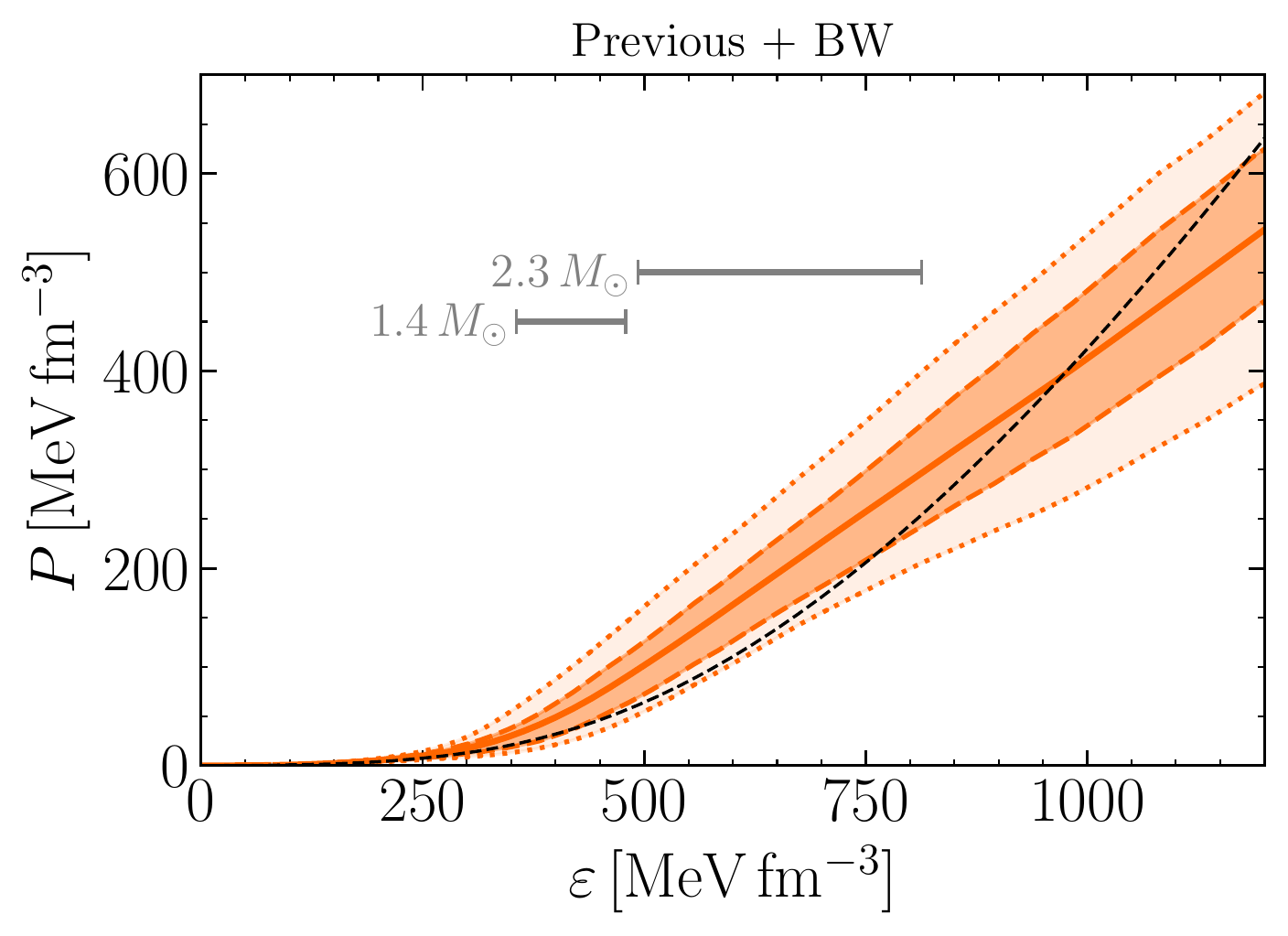} \\
		\includegraphics[height=55mm,angle=-00]{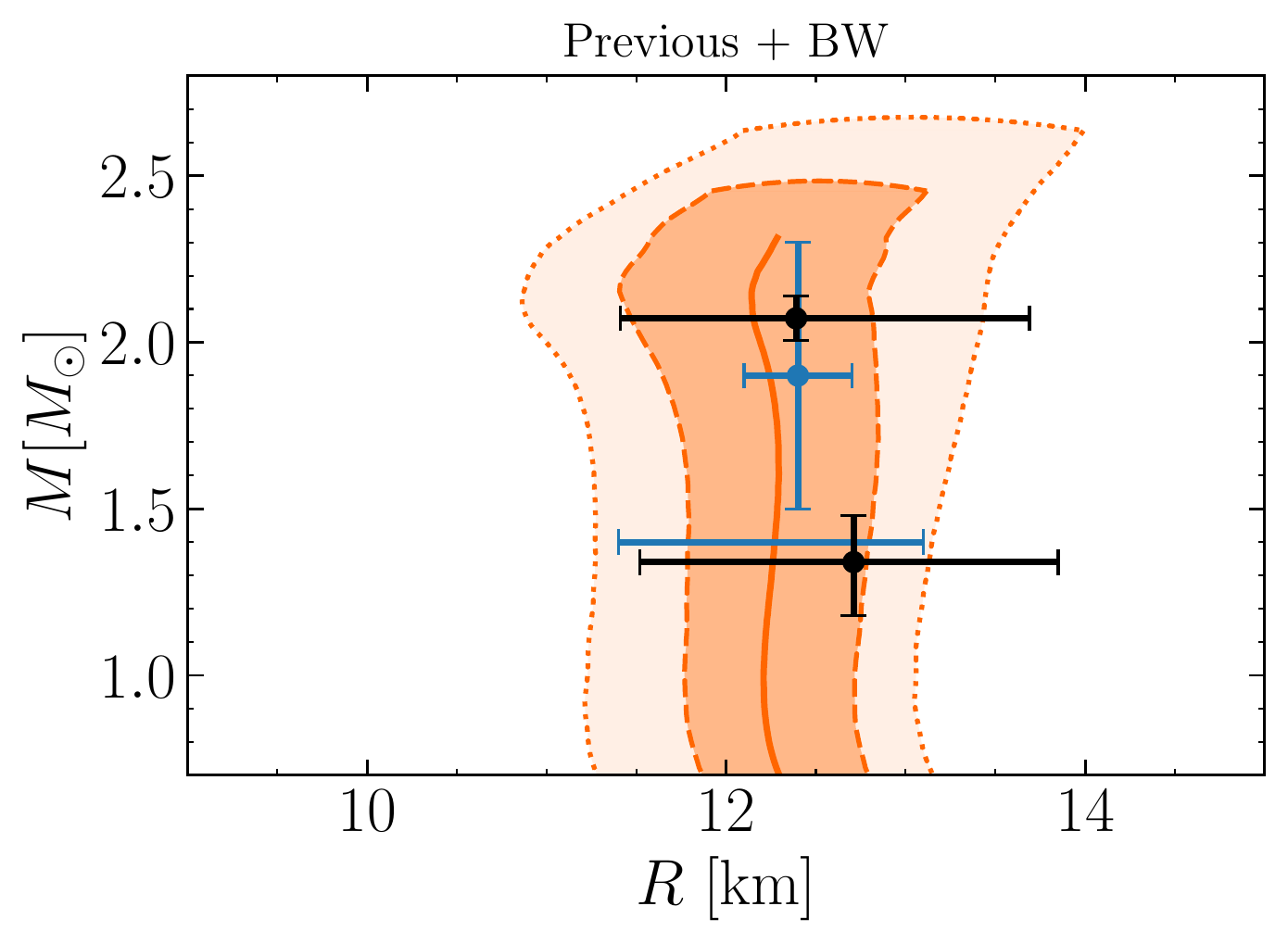} 
		\includegraphics[height=55mm,angle=-00]{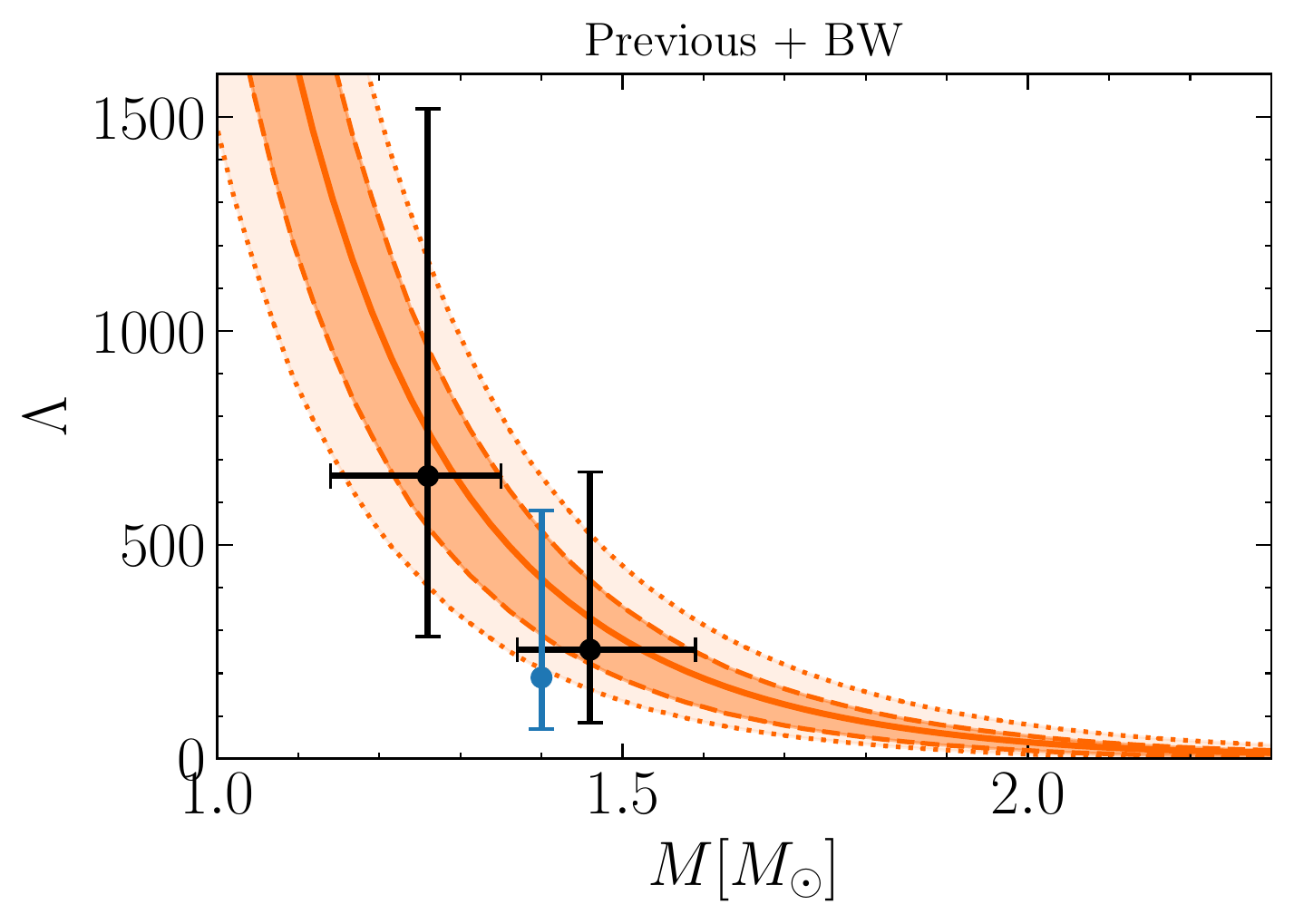}
		\caption{Marginal posterior probability distributions at the 95\% and 68\% levels: squared speed of sound $c_s^2$ and pressure $P$ as a function of energy density $\varepsilon$, inferred from the dataset 'Previous + BW' (see Tab.\,\,\ref{tab:DataSet}). Also shown are the marginal posterior probability distributions for the mass-radius relation and the tidal deformability, $\Lambda$, as a function of neutron star mass $M$ in units of the solar mass $M_\odot$. At each $\varepsilon$ or $M$, there exist 95\% and 68\% posterior credible intervals for $c_s^2(\varepsilon)$, $P(\varepsilon)$ or $R(M)$, $\Lambda(M)$. These intervals are connected to obtain the posterior credible bands. Similarly, the medians of the posterior probability distributions at each $\varepsilon$ or $M$ are connected (solid lines). For squared speed of sound or pressure the dashed black lines indicate the value of the conformal limit or represent the APR EoS \cite{Akmal1998}.  In the upper two figures bars mark the 68\% credible intervals of the central energy densities of neutron stars with masses $M = 1.4 \, M_\odot$ and $2.3 \, M_\odot$, respectively. The mass-radius relation is compared to the marginalized intervals at the 68\% level from the NICER data analyses by Riley \textit{et al.} (black) \cite{Riley2019,Riley2021} of PSR J0030+0451 and PSR J0740+6620.  In addition the 68\% mass-radius credible intervals of the thermonuclear burster 4U 1702-429 \cite{Naettilae2017} are displayed as well as the 68\% credible interval of $R(1.4\,M_\odot)$ extracted from quiescent low-mass x-ray binaries \cite{Steiner2018} (blue), both of which are not included in the Bayesian analysis. $\Lambda(M)$ is compared to the masses and tidal deformabilities inferred in Ref.\,\,\cite{Fasano2019} for the two neutron stars in the merger event GW170817 at the 90\% level (black) as well as $\Lambda(1.4\,M_\odot)$ at the 90\% level extracted from GW170817  \cite{Abbott2018} (blue).}
		\label{fig:PosteriorBandsBW}
	\end{center}
\end{figure*}

The posterior credible bands for the mass-radius relation are displayed in Fig.\,\,\ref{fig:PosteriorBandsBW}. Median and credible bands terminate at the respective median and upper limits of $M_{\max}$. The posterior mass-radius credible bands are in good agreement with the 68\% credible intervals inferred from the NICER measurements of PSR J0030+0451 and PSR J0740+6620. The $R(M)$ credible band is slightly shifted to smaller radii compared to the NICER data because the gravitational wave event GW170817 prefers such smaller radii \cite{Raithel2018}. The 68\% credible interval for the radius of a neutron star with mass $M = 1.4\,M_\odot$ agrees well with the radius extracted from quiescent low-mass x-ray binaries in the baseline scenario of Ref.\,\,\cite{Steiner2018}. Moreover, the 68\% mass and radius credible intervals of 4U 1702-429 extracted from thermonuclear x-ray bursts in Ref.\,\,\cite{Naettilae2017} lie well within the 68\% mass-radius credible band.  Notably both of these observations were not used as input in the Bayesian inference procedure.  We mention that in Ref.\,\,\cite{AlMamun2021} quiescent low-mass x-ray binaries and sources of thermonuclear bursts are also found to fit into an overall picture that is consistent with the gravitational wave and NICER data. 

The 90\% credible intervals for the tidal deformabilities and masses of the two neutron stars in the merger event GW170817 extracted in Ref.\,\,\cite{Fasano2019} agree well with the posterior credible band of $\Lambda(M)$ displayed in Fig.\,\,\ref{fig:PosteriorBandsBW}. Using universal relations the tidal deformability of a $1.4\,M_\odot$ neutron star was extracted based on the information from the merger event GW170817 in Ref.\,\,\cite{Abbott2018}. The resulting 90\% credible interval does agree with the posterior credible bands of $\Lambda(M)$ at $M = 1.4\,M_\odot$, which lie, however, at slightly larger tidal deformabilities.

\begin{figure}[tbp]
	\begin{center}
		\includegraphics[height=55mm,angle=-00]{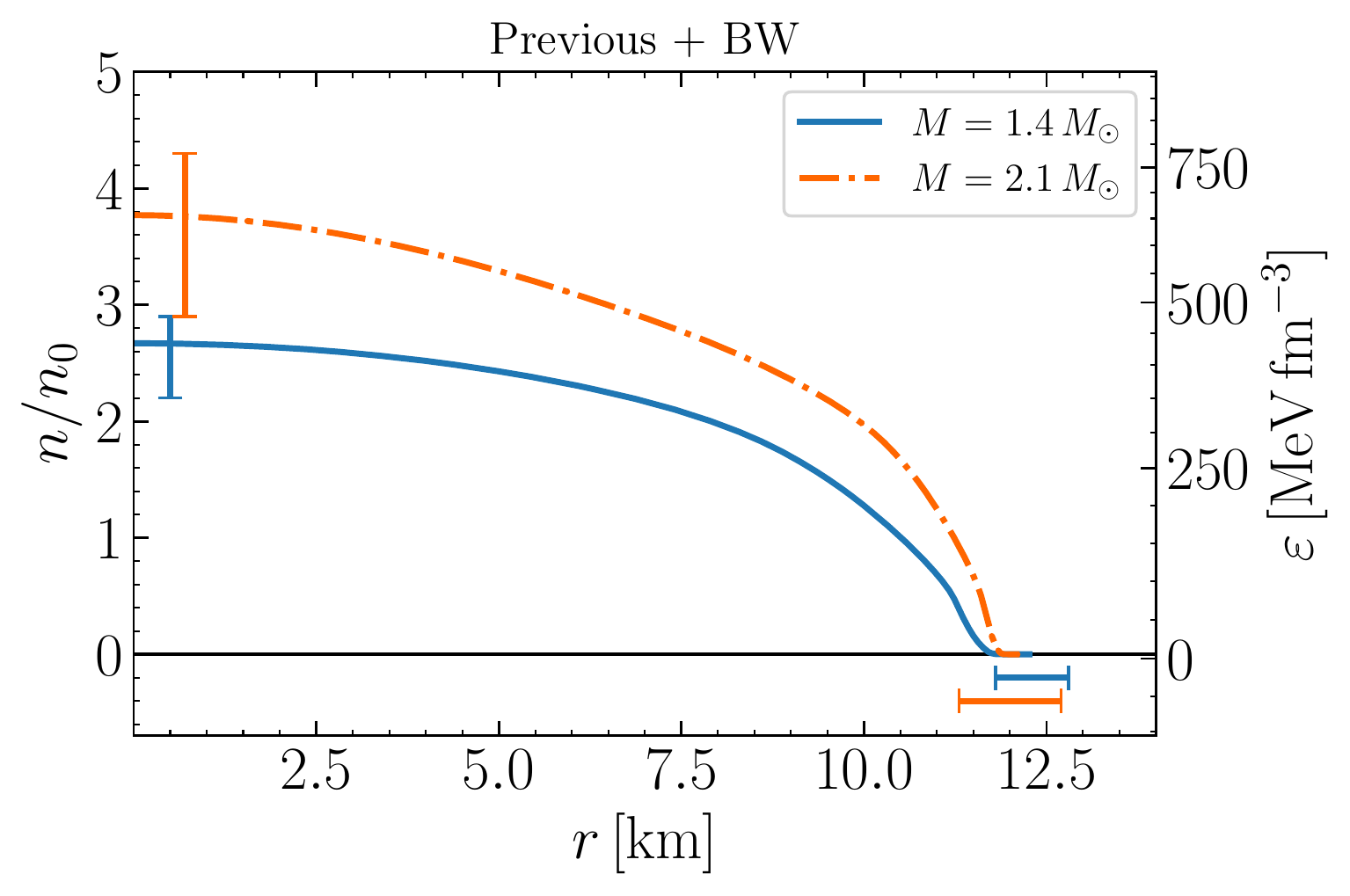}
		\caption{Density profiles of neutron stars with masses of $M=1.4\,M_\odot$ and $M=2.1\,M_\odot$. The employed equation of state corresponds to the median of the credible band in Fig.\,\,\ref{fig:PosteriorBandsBW}, i.e. using the previously available and the new data from the black widow pulsar. The bars indicate the 68\% credible intervals of the central densities and radii of neutron stars with mass $M = 1.4\,M_\odot$ (blue) and $2.1\,M_\odot$ (orange),  as listed in Tab.\,\,\ref{tab:NS_properties1}.}
		\label{fig:DensityProfiles}
	\end{center}
\end{figure}

To give an impression of the matter distribution inside neutron stars,  the density profiles of objects with masses  $M=1.4\,M_\odot$ and $M=2.1\,M_\odot$ are displayed in Fig.\,\,\ref{fig:DensityProfiles}. These profiles are computed using the median of $P(\varepsilon)$ based on the previously available data together with the new information from the black widow pulsar. On the axes of Fig.\,\,\ref{fig:DensityProfiles} the 68\% credible intervals of the central densities and radii of $1.4$ and $2.1\,M_\odot$ neutron stars listed in Tab.\,\,\ref{tab:NS_properties1} are indicated for comparison.  The skewness of the posterior probability distribution makes the central densities and radii in Fig.\,\,\ref{fig:DensityProfiles} deviate slightly from the median values listed in Tab.\,\,\ref{tab:NS_properties1}.  As in Fig.\,\,\ref{fig:PosteriorBandsBW} both neutron stars with masses $M=1.4\,M_\odot$ and $2.1\,M_\odot$ have almost equal radii. The density profiles smoothly decrease towards small densities in the outer regions of the stars.  In this regime the EoS is governed by the ChEFT constraint.

\subsection{Speed of sound and EoS}

\begin{figure*}[tbp]
	\begin{center}
		\includegraphics[height=55mm,angle=-00]{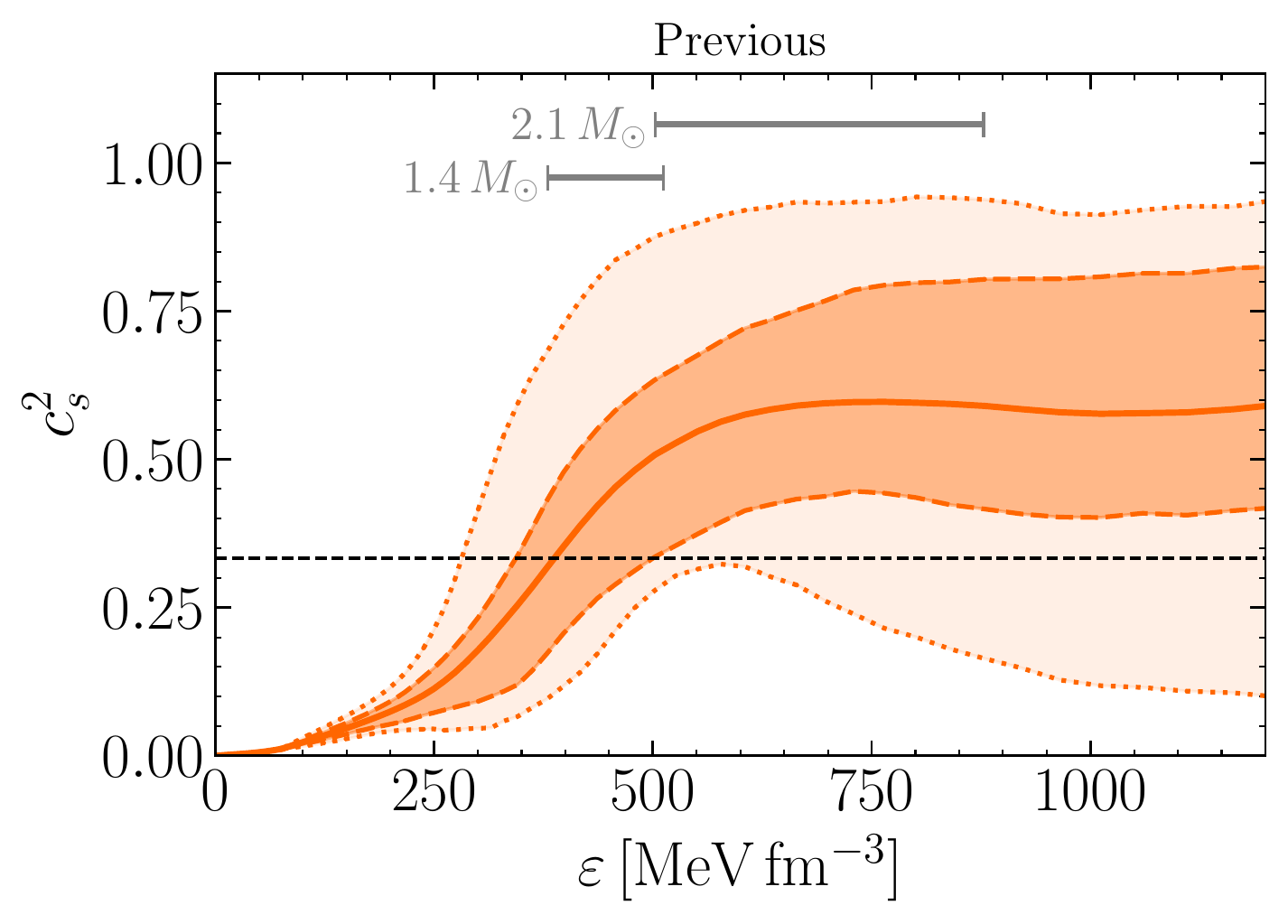} 
		\includegraphics[height=55mm,angle=-00]{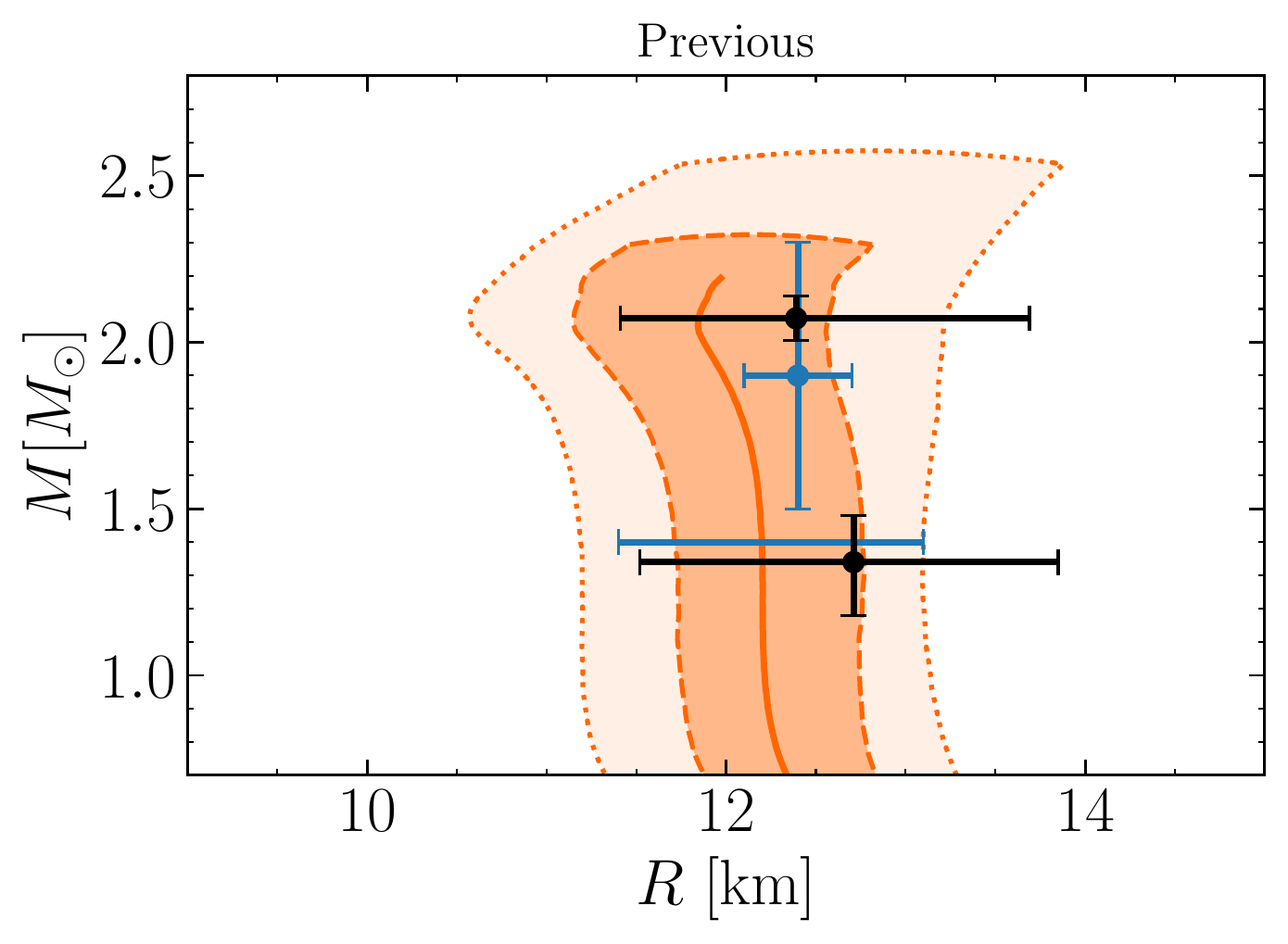} 
		\caption{Similar to Fig.\,\,\ref{fig:PosteriorBandsBW}: posterior credible bands are displayed for the squared speed of sound, $c_s^2$,  as a function of energy density $\varepsilon$,  and the mass-radius relation $R(M)$,  but now using only the 'Previous' data in Tab.\,\,\ref{tab:DataSet} without inclusion of the new PSR J0952-0607 (BW) information.}
		\label{fig:PosteriorBandsPrev}
	\end{center}
\end{figure*}

The posterior credible bands of the speed of sound and the pressure as a function of energy density,  based on the previously available data plus the new superheavy mass observation,  are displayed in Fig.\,\,\ref{fig:PosteriorBandsBW}. With the updated ChEFT likelihood of Eq.\,(\ref{eq:LikelihoodChEFT}), the sound velocity remains small at low energy densities,  $\varepsilon < 250\,$MeV$\,$fm$^{-3}$. Given that the low-density behaviour is constrained up to $n_\text{ChEFT} = 1.3\,n_0$ (unlike some previous setups that used $n_\text{ChEFT} = 2\,n_0$),  a steeper increase of $c_s^2$ is now possible at densities around twice $n_0$,  i.e.  from energy densities $\varepsilon \sim 250 - 300\,$MeV$\,$fm$^{-3}$ onward. 

The median of $c_s^2(\varepsilon)$ exceeds the conformal limit, $c_s^2 = 1/3$, around $\varepsilon \sim 350\,$MeV$\,$fm$^{-3}$. We can quantify the evidence for a violation of the conformal bound by computing the Bayes factor $\mathcal{B}^{c_{s,\text{max}}^2 > 1/3}_{c_{s,\text{max}}^2 \leq 1/3}$ which compares equations of state with maximum squared speed of sound larger than $1/3$ to EoSs with maximum squared sound speed below  $1/3$. With a Bayes factor of well over $10^3$, there is extreme evidence that $c_s^2$ exceeds the conformal bound inside neutron stars.  This is consistent with other recent studies \cite{Landry2019,Landry2020,Legred2021,Leonhardt2020,Altiparmak2022,Gorda2023, Brandes2023}. 

At intermediate energy densities, $\varepsilon \sim 600\,$MeV$\,$fm$^{-3}$, the speed of sound starts to form a  plateau that extends up to higher energy densities. This behaviour of the sound speed is reflected in the pressure.  The plateau in $c_s^2$ corresponds to an approximately linear rise of the pressure with increasing energy density.   In a double logarithmic depiction of $P(\varepsilon)$ the onset of this behaviour is reminiscent of the 'kink' noted in Ref.\,\,\cite{Annala2020}.  It is then apparent that such a 'kink' is not necessarily a signal of a pronounced softening of the EoS but may just reflect the formation of a plateau in the squared sound velocity.

At energy densities $\varepsilon \sim 350 - 950\,$MeV$\,$fm$^{-3}$, the inferred equation of state turns out to be stiffer than the Akmal-Pandharipande-Ravenhall (APR) EoS \cite{Akmal1998}, whereas at higher energy densities $P(\varepsilon)$ increases more slowly as compared to APR.  (We recall the known feature that the APR equation of state violates causality at the highest densities.)

In Fig.\,\,\ref{fig:PosteriorBandsPrev} the posterior credible bands of the sound speed and mass-radius relation are displayed for an inference without the new information from PSR J0952-0607.  At first sight the comparison between Figs.\,\,\ref{fig:PosteriorBandsPrev} and \ref{fig:PosteriorBandsBW} appears to reveal only marginal differences.  But with a more focused view the absence of the condition to reach the high mass of the black widow pulsar implies that the speed of sound increases more softly at low energy densities, $\varepsilon \lesssim 500\,$MeV$\,$fm$^{-3}$.  Accordingly,  the conformal bound is exceeded at higher energy densities, and smaller speeds of sound are reached compared to the case including the heavy mass data in Fig.\,\,\ref{fig:PosteriorBandsBW}.  Without inclusion of the new data smaller radii are possible for heavy neutron stars with $M \sim 2.1\,M_\odot$.  In contrast, the mass-radius relation in Fig.\,\,\ref{fig:PosteriorBandsBW} features an almost constant radius over the whole mass range.

\subsection{Evidence for (or against) a strong\\ first-order phase transition}

\begin{figure}[tbp]
	\begin{center}
		\includegraphics[height=55mm,angle=-00]{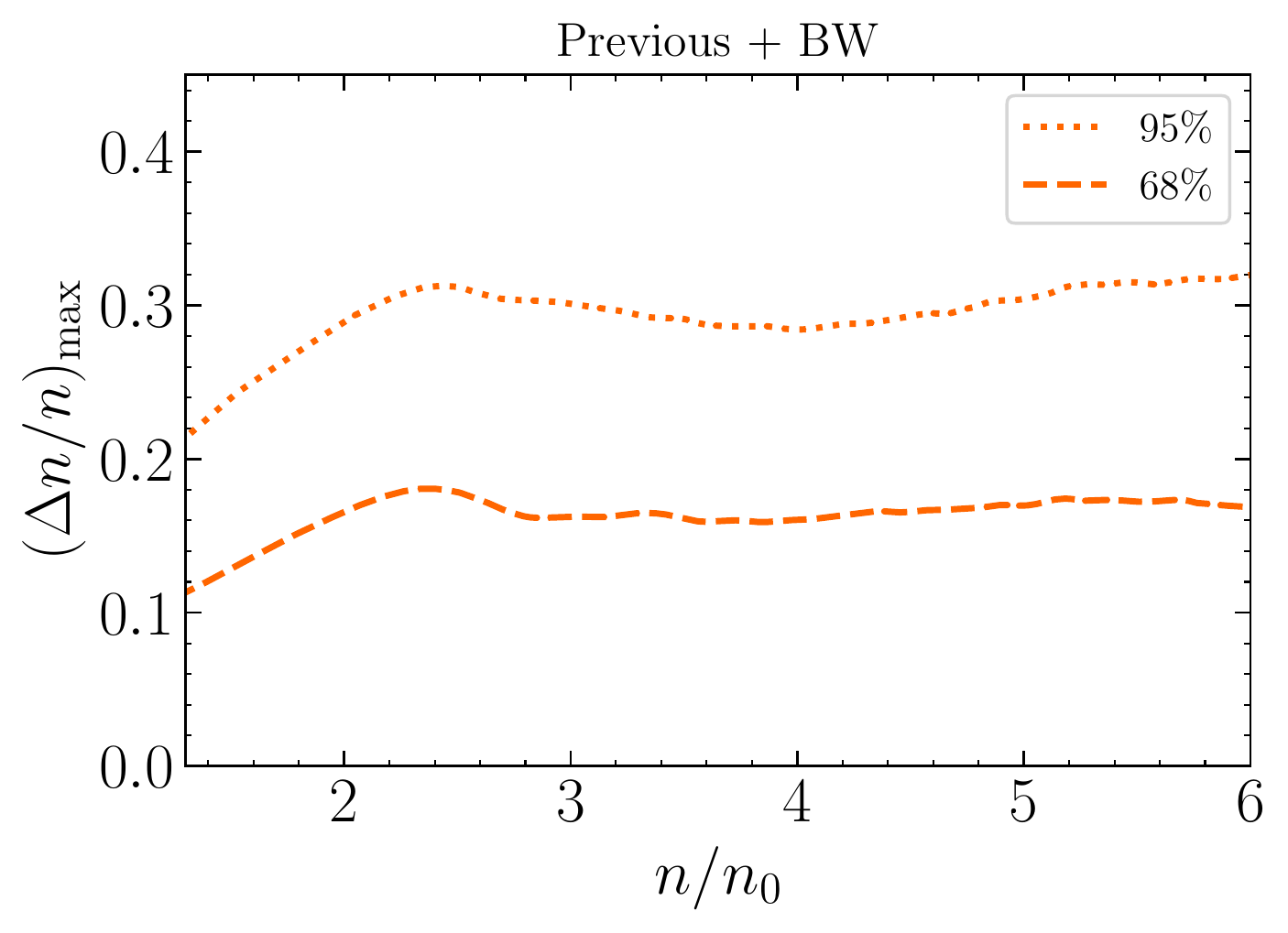}
		\caption{Maximum possible phase coexistence interval $(\Delta n/n)_{\max}$ of constant pressure (where $n$ is the density at which the interval starts) extracted from the 68\% and 95\% posterior credible bands of $P(n)$. $(\Delta n/n)_{\max}$ is displayed as a function of baryon density $n$ in units of the nuclear saturation density, $n_0 = 0.16\,$fm$^{-3}$.}
		\label{fig:CoexistenceInterval}
	\end{center}
\end{figure}

A sufficient condition for a first-order phase transition is an equation-of-state that features a domain of phase coexistence within which a Maxwell construction implies a region of constant pressure.  The width of this domain,  characterized by $\Delta n/n$ (where $n$ is the density at which the coexistence interval starts),  is a measure of the ‘strength’ of the phase transition.  For guidance and comparison,  an example of a ‘strong’ first-order transition is the liquid-gas phase transition in symmetric nuclear matter. There,  at low temperatures ($T < 15\,$MeV),  the phase coexistence region obtained through a Maxwell construction has a typical width $\Delta n/n > 1$ \cite{Fiorilla2012,Wellenhofer2014,Brandes2021}.

Starting from a given EoS,  $P(\varepsilon)$,  the Gibbs-Duhem relation is used to re-express pressure as a function of density, $P(n)$. 
The border lines of the 68\% and 95\% posterior credible bands for $P(n)$ constrain the maximum possible phase coexistence intervals,  $(\Delta n/n)_{\max}$,  at the corresponding credibility levels.  The results in Fig.\,\,\ref{fig:CoexistenceInterval} show that these maximum possible coexistence regions are narrow: $(\Delta n/n)_{\max} \simeq 0.2$ $(0.3)$ at the 68\% (95\%) level.  In fact it turns out that $(\Delta n/n)_{\max}$ is nearly constant as a function of the baryon density $n$ taken at the starting point of the possible phase coexistence region.  This observation holds throughout the regime relevant for neutron stars.  We conclude that only weak phase transitions with $\Delta n/n \leq (\Delta n/n)_{\max}$ can still be realized inside neutron stars within the inferred posterior credible bands.

The detailed behaviour of the squared speed of sound is to be seen in a related context. Figs.\,\,\ref{fig:PosteriorBandsBW} and \,\,\ref{fig:PosteriorBandsPrev} show indications of a shallow maximum, $c_{s,\max}^2$.  At the 68\% level this maximum takes a value $c_{s,\max}^2 = 0.78_{-0.11}^{+0.18}$ at a baryon density $n(c_{s,\max}^2) = 3.2_{-1.2}^{+0.8}\,n_0$.  The peak in $c_s^2$ found in Ref.\,\,\cite{Marczenko2023} has a similar magnitude and location,  although a pronounced peak structure is not seen in our posterior result.  In contrast we find that the sound velocity forms a plateau at higher densities.  There is no indication of a softening.  Still,  at the 95\% level small sound speeds are not entirely excluded,  though the probability of their occurrence is low.  Nevertheless,  at asymptotically high densities pQCD dictates that the speed of sound reaches the conformal bound $c_s^2 = 1/3$ from below.  This implies that at some density beyond the plateau,  the speed of sound  must turn downward again and reach a minimum,  $c_{s,\min}^2$,  at some higher density.  A fast drop in $c_s^2$ could potentially indicate the occurrence of a phase transition.  The question is whether such a decrease still takes place within the density range of neutron star cores.  

\begin{figure}[tbp]
	\begin{center}
		\includegraphics[width=85mm,angle=-00]{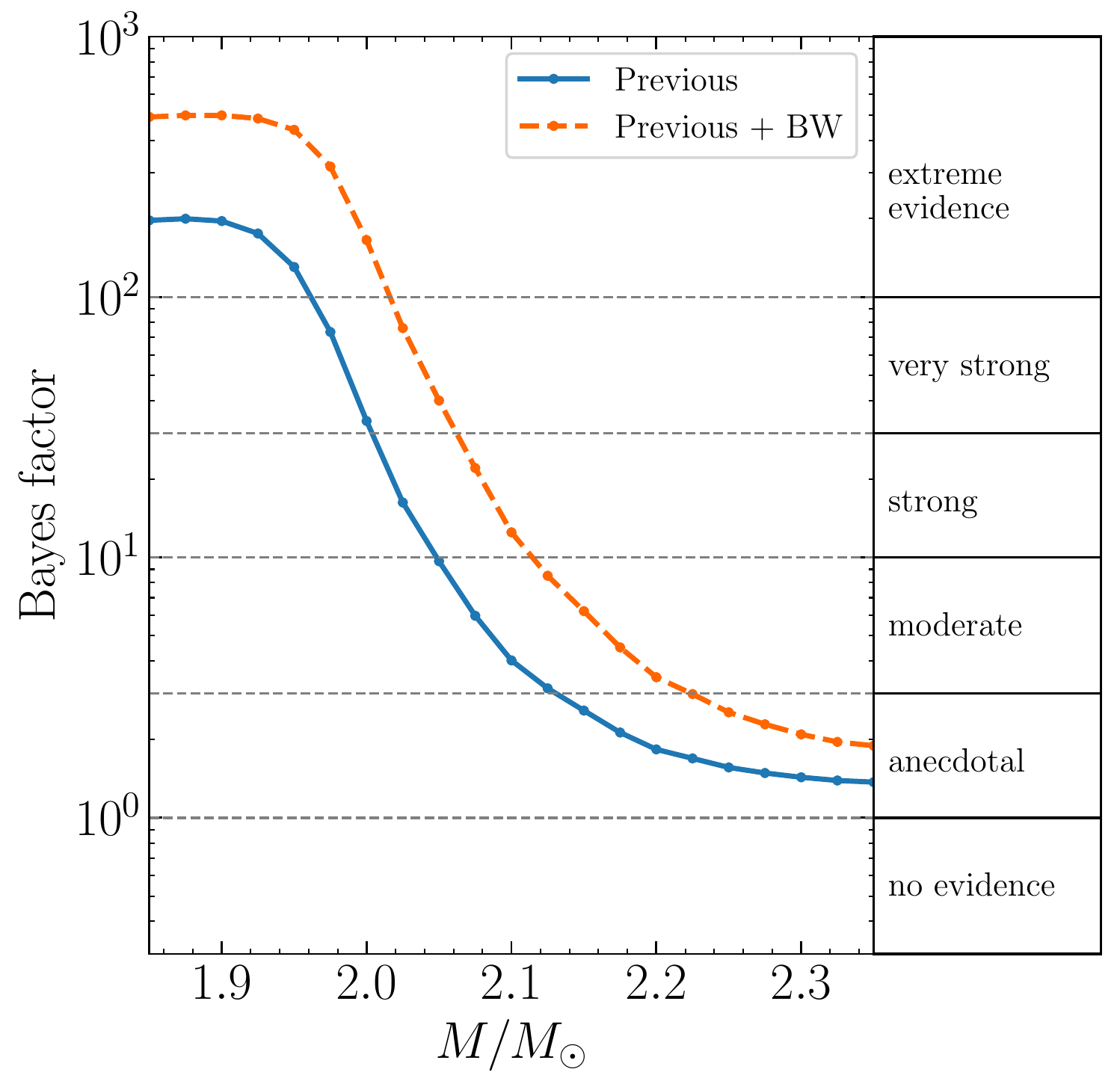} \\
		\caption{Bayes factors $\mathcal{B}^{c_{s,\min}^2 > 0.1}_{c_{s,\min}^2 \leq 0.1}$ comparing EoS samples with the following competing scenarios: a) minimum squared speed of sound  (following a maximum),  with $c^2_{s,\min}$ larger than 0.1,  excluding a strong first-order phase transition with a Maxwell construction; versus b) EoS samples with $c_{s,\min}^2 \leq 0.1$.  The Bayes factors are calculated for a given maximum neutron star mass $M$,  i.e. the minimum speed of sound up to the corresponding mass is used.  For illustration the evidence classification from Refs.\,\,\cite{Jeffreys1961,Lee2016} is indicated by dashed grey lines and annotated on the right hand side. }
		\label{fig:BayesFactor}
	\end{center}
\end{figure}

To answer this question we specifically perform a Bayes factor analysis to quantify the evidence for a rapid variation of energy density with pressure,
corresponding to a low averaged sound speed over the relevant pressure
interval. With this aim Bayes factors $\mathcal{B}^{c_{s,\min}^2 > 0.1}_{c_{s,\min}^2 \leq 0.1}$ are computed,  comparing the evidence for EoSs with a minimum speed of sound larger than $c_{s,\min}^2 > 0.1$ over EoSs with small $c_{s,\min}^2 \le 0.1$,  the latter possibly indicating a strong first-order phase transition.  It is assumed that this minimum is positioned above the maximum located at lower densities, $n(c_{s,\min}^2) > n(c_{s,\max}^2)$.  These Bayes factors are shown in Fig.\,\,\ref{fig:BayesFactor},  calculated for a given maximum mass,  i.e.  the minimum speed of sound up to the corresponding $M$ is used in the likelihood computation.  As in Ref.\,\,\cite{Brandes2023} there is extreme or very strong evidence against small sound speeds inside neutron stars with masses up to $M\leq 2\, M_\odot$. The Bayes factors increase further with the inclusion of the black widow (BW) pulsar information.  With these new data,  there is strong evidence against small sound speeds, $c_{s,\min}^2 <0.1$,  inside neutron stars with masses even up to $M \leq 2.1 \, M_\odot$.

 In Refs.\,\,\cite{Ecker2022a,Annala2023}, the authors also find sound speeds larger than $c_s^2 > 0.1$ at the 95\% level for neutron stars with mass $M = 2\,M_\odot$.  In their analyses the authors use different parametrizations and accordingly different prior distributions. With that much consistent evidence it is safe to say that a strong first-order phase transition in the core of neutron stars with mass $M \leq 2.0\, M_\odot$ is fairly unlikely based on the current data.  With the new information from the black widow pulsar PSR J0952-0607 the Bayes factors in Fig.\,\,\ref{fig:BayesFactor} further increase such that the evidence against small sound speeds inside even heavier neutron stars with masses up to $M\leq 2.1\, M_\odot$ becomes strong. The Bayes factors feature a plateau at extreme evidence for maximum masses smaller than $M \lesssim 1.9\,M_\odot$,  because all relevant EoSs must support these masses in order to fulfil the Shapiro and NICER constraints. Numerical values of the Bayes factors $\mathcal{B}^{c_{s,\min}^2 > 0.1}_{c_{s,\min}^2 \leq 0.1}$ for different maximum masses can be found in Tab.\,\,\ref{tab:BayesFactorSmallCsM} in Appendix\,\ref{sec:BayesFactorTable}.  The Bayes factors corresponding to a stronger criterion of smaller minimum speeds of sound, $\mathcal{B}^{c_{s,\min}^2 > 0.05}_{c_{s,\min}^2 \leq 0.05}$, are larger especially at small maximum masses, but lead to similar evidence classifications. \\

It turns out that just four segments are sufficient to cover the entire set of astrophysical data and theory constraints,  in agreement with the findings of Refs.\,\,\cite{Altiparmak2022,Annala2023}.  With up to three more segments available,  the parametrization employed in the present work has sufficient flexibility to describe additional features such as phase transitions,  if these can occur within the given range of uncertainties.  We actually find similar Bayes factors compared to our previous work in Ref.\,\,\cite{Brandes2023},  where less general parametrizations have been used.  This indicates that our results are not influenced by details of the parametrization.

In addition to setting constraints for a strong first-order phase transition in the core of neutron stars, the Bayes factors in Fig.\,\,\ref{fig:BayesFactor} also limit the likelihood for the appearance of a continuous crossover with $c_s^2 \leq 0.1$. A softer crossover with $c_s^2 > 0.1$ is still possible in neutron stars with masses up to $M\lesssim 2.1\,M_\odot$ and beyond.  This includes EoSs featuring quark-hadron continuity \cite{Baym2018,Kojo2022} or percolation scenarios \cite{Fukushima2020}.  Moreover,  small sound speeds $c_s^2 < 0.1$ in the cores of neutron stars with even higher masses,  $M\gtrsim 2.2\,M_\odot$,  less constrained by the currently available astrophysical data,  cannot be firmly excluded.  Similarly,  a phase transition with a Gibbs (rather than a Maxwell) construction \cite{Han2019} does not necessarily result in a drop of the sound speed to $c_s^2 \sim 0$ and can also not be ruled out.

By analysing minima occuring at densities beyond the maximum of the speed of sound,  $n(c_{s,\min}^2) > n(c_{s,\max}^2)$,  we are restricting ourselves to setting constraints for strong first-order phase transitions in the deep core of neutron stars.  From the behaviour of the sound speed in Fig.\,\,\ref{fig:PosteriorBandsBW},  a minimum appearing at a density lower than that of the maximum seems to be only conceivable at small energy densities, $\varepsilon \sim 250 - 350\,$MeV$\,$fm$^{-3}$,  i.e.  at baryon densities $n\lesssim 2\,n_0$. With this in mind,  we proceed in the next section to quantify the evidence for the possible occurrence of a phase transition strong enough to lead to a disconnected mass-radius relation. 

\subsection{Twin-star scenarios}

Among the multitude of possible equations of state in the prior,  3.5\% have a disconnected mass-radius relation with more than one stable branch and hence represent a possible twin-star scenario.  To quantify the evidence for such a scenario,  we compute Bayes factors $\mathcal{B}_{N_\text{branches} > 1}^{N_\text{branches} = 1}$ comparing the marginalized likelihoods of EoSs with a single connected mass-radius relation to EoSs with multiple stable branches.  The resulting Bayes factor of well over 900 demonstrates extreme evidence against a disconnected mass-radius relation with multiple stable branches. This value further increases with the inclusion of the new data from the black widow pulsar.  The conclusion agrees well with that of Ref.\,\,\cite{Gorda2022a} where the authors find only an extremely small possible parameter space for a twin-star scenario that is consistent with the low-density constraint from ChEFT and the astrophysical data. Furthermore, the authors already note that the observation of a still more massive neutron star beyond $M \simeq 2\,M_\odot$ would make a twin-star scenario even more unlikely. 

If the low-density constraint involving likelihood from ChEFT is ignored, the pertinent Bayes factor decreases to $\mathcal{B}_{N_\text{branches} > 1}^{N_\text{branches} = 1} = 11.8$, providing 'only' strong evidence against a scenario with multiple disconnected branches. This value increases to $\mathcal{B}_{N_\text{branches} > 1}^{N_\text{branches} = 1} = 13.0$ with the inclusion of the new heavy-mass data.  In comparison, the Bayesian analyses in Refs.\,\,\cite{Legred2021,Essick2023},  where the authors do not employ ChEFT information find only moderate evidence. This difference may be traced to the different treatment of the neutron star crust.  It appears that the only possibility for a twin-star scenario,  given the astrophysical data base as a constraint,  is through a phase transition that takes place at very low energy densities shortly above those in the neutron star crust,  which is also noted in Ref.\,\,\cite{Essick2023}.  Accordingly,  the mass at which the mass-radius branches become disconnected is as low as $M \sim 0.8\,M_\odot$.  (Note that in our analysis similar to Ref.\,\,\cite{Essick2023}, we do not consider disconnected branches below the assumed minimum neutron star mass, $M_{\min} = 0.5\, M_\odot$.)

The additional inclusion of the new information from HESS J1731-347 further strengthens the evidence against a twin-star scenario,  even in the absence of ChEFT constraints: the evidence now becomes very strong with a Bayes factor of $\mathcal{B}_{N_\text{branches} > 1}^{N_\text{branches} = 1} \simeq 35$. This is quite interesting as some authors considered the unusually light HESS supernova remnant as a hint in favour of a twin-star scenario \cite{Tsaloukidis2023}. 

\subsection{Trace anomaly measure}

\begin{figure}[tbp]
	\begin{center}
		\includegraphics[height=55mm,angle=-00]{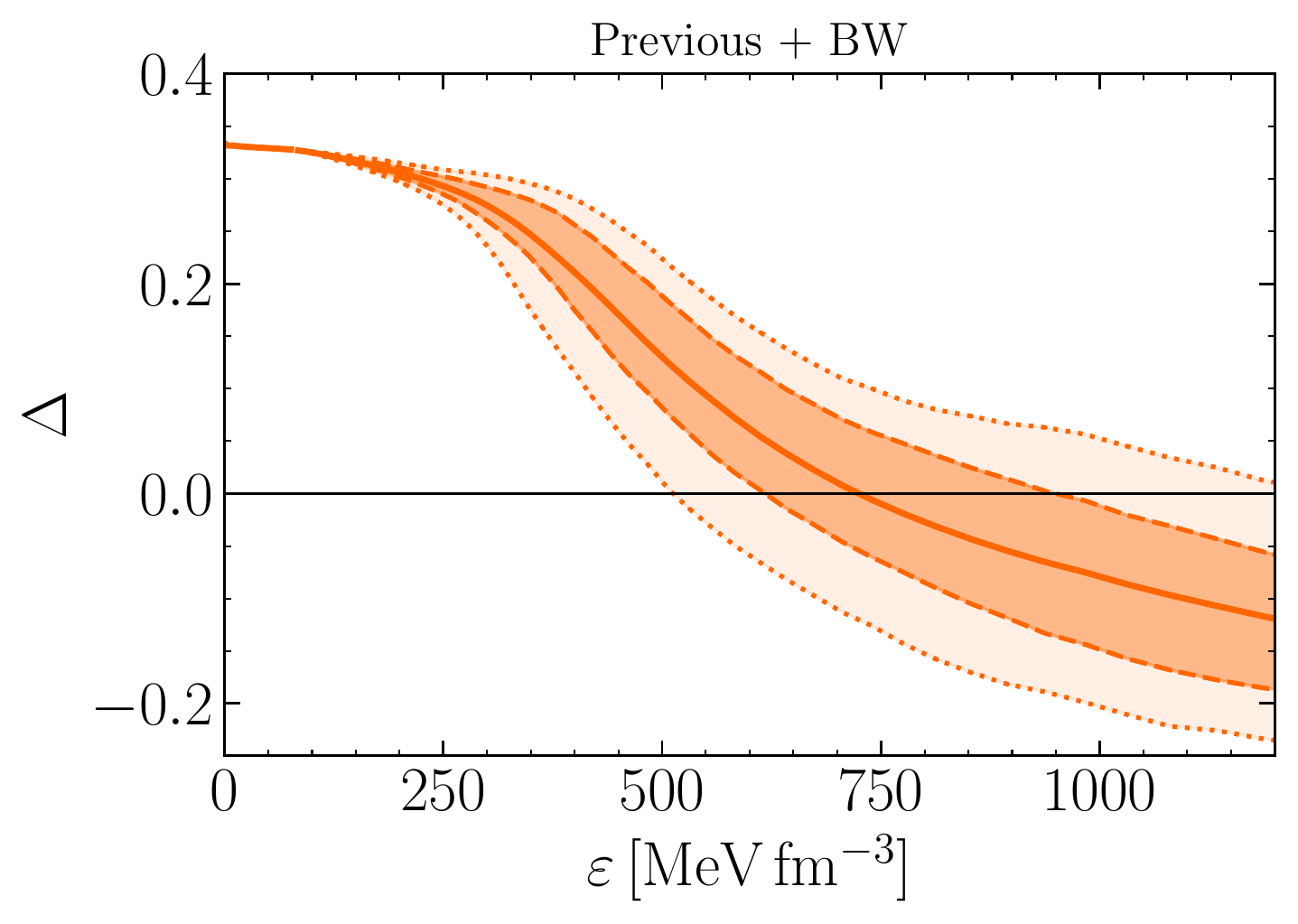}
		\caption{Posterior 95\% and 68\% credible bands and medians for the trace anomaly measure $\Delta = 1/3 - P/\varepsilon$ as a function of energy density $\varepsilon$.}
		\label{fig:Delta}
	\end{center}
\end{figure}

Based on the equation of state $P(\varepsilon)$ the trace anomaly measure $\Delta$ can be computed,  given by the normalized trace of the energy momentum tensor $T^{\mu\nu}$:
\begin{equation}
	\Delta = \frac{g_{\mu\nu}T^{\mu\nu}}{3\varepsilon} = \frac{1}{3} - \frac{P}{\varepsilon}\,.
\end{equation}
Causality and thermodynamic stability dictate that the trace anomaly measure has to be within the range $-2/3 \geq \Delta \geq 1/3$. Moreover,  $\Delta\rightarrow 0$ for conformal matter realized at high densities. The posterior credible bands for the trace anomaly measure are shown in Fig.\,\,\ref{fig:Delta}. Starting with a value $\Delta = 1/3$ at zero density,  the trace anomaly measure decreases with increasing energy density until $\varepsilon \sim 700\,$MeV$\,$fm$^{-3}$, where the median of $\Delta$ turns negative.  At even higher energy densities, $\varepsilon \gtrsim 900\,$MeV$\,$fm$^{-3}$,  encountered only in extremely heavy neutron stars,  the 68\% credible band becomes altogether negative.  

In order to access the evidence for a negative trace anomaly measure we compute Bayes factors $\mathcal{B}_{\Delta \geq 0}^{\Delta < 0}$,  comparing the likelihood for EoSs with negative trace anomaly, $\Delta < 0$,  up to $\varepsilon_{c,\max}$ versus EoSs with positive $\Delta$. Given only the previously available data with a resulting Bayes factor $\mathcal{B}_{\Delta \geq 0}^{\Delta < 0} = 6.32$,  there is moderate evidence that $\Delta$ becomes negative within neutron stars. The Bayes factor further increases to $\mathcal{B}_{\Delta \geq 0}^{\Delta < 0} = 8.11$ with the inclusion of the new information from PSR J0952-0607.  These results are consistent with the deduced empirical band for $\Delta$ in Refs.\,\,\cite{Fujimoto2020,Fujimoto2022a},  which also starts turning negative around $\varepsilon \sim 700\,$MeV$\,$fm$^{-3}$. At the same time, 
the authors of Ref.\,\,\cite{Fujimoto2022a} motivate a scenario with positive trace anomaly measure $\Delta \geq 0$, which is in light contrast to our Bayes factor analysis.    

Lattice QCD calculations suggest that the trace anomaly measure always stays larger than zero at finite temperatures and vanishing baryon chemical potential \cite{Borsanyi2014,Bazavov2014}.  However,  in two-colour QCD the trace anomaly can become negative at finite chemical potentials \cite{Cotter2013,Iida2022}. Other recent Bayesian studies have also found a negative trace anomaly measure at high densities \cite{Marczenko2023,Takatsy2023,Annala2023} or in extremely heavy neutron stars \cite{Ecker2022}.  At much higher energy densities beyond those displayed in Fig.\,\,\ref{fig:Delta},  the asymptotic pQCD limit does imply a switch back to positive $\Delta$ in the approach to $\Delta \rightarrow 0$.

\subsection{Impact of low-density (ChEFT) constraint}

The low-density conditioning of the EoS must incorporate the breadth of well-known empirical facts from nuclear physics.  Chiral effective field theory is an established framework for this purpose.  We recall that we are taking a conservative position here,  terminating the applicability range of ChEFT at $ n_\text{ChEFT} = 1.3\, n_0$.  

\begin{figure}[tbp]
	\begin{center}
		\includegraphics[height=55mm,angle=-00]{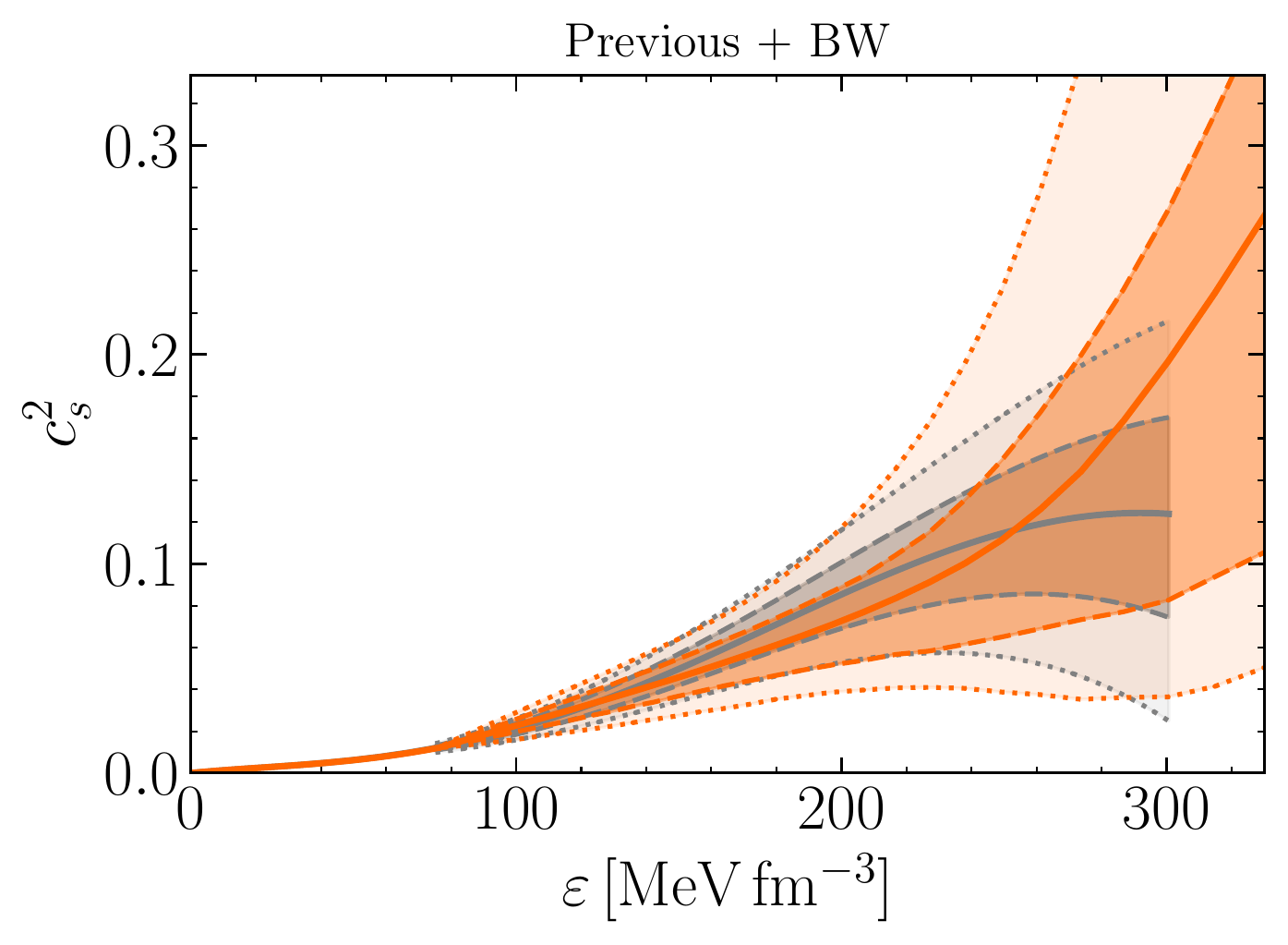}
		\caption{Posterior credible bands for the squared speed of sound, $c_s^2$,  as a function of energy density $\varepsilon$.  The inference includes previous data as well as the new information from the black-widow pulsar (BW). The low density behaviour of the posterior credible bands is compared to the N3LO ChEFT results from Refs.\,\,\cite{Drischler2021,Drischler2022} (grey) up to 300$\,$MeV$\,$fm$^{-3}$. }
		\label{fig:PosteriorBandsChEFT}
	\end{center}
\end{figure}

The ChEFT constraints do indeed provide an important limiting window for the evolution of the EoS to higher densities.  It is thus of interest to analyse the impact of the ChEFT likelihood in the inference procedure.  We do this by comparing the posterior credible bands of the speed of sound as they emerge in our approach,  to the ChEFT constrained results from Refs.\,\,\cite{Drischler2021,Drischler2022} which extend up to $n_\text{ChEFT} = 2.0\,n_0$ (see Figure \ref{fig:PosteriorBandsChEFT}). Two important findings concerning the ChEFT impact in relation to constraints from astrophysical data become apparent from this figure.  First,  at small energy densities the $c_s^2$ posterior has extra support at small sound speeds below the ChEFT constraint,  because small $c_s^2$ are preferred by the gravitational wave event GW170817.  Lowering the ChEFT constraint density to $n_\text{ChEFT} = 1.1\,n_0$ therefore changes the posterior credible bands only marginally.  However,  at energy densities $n >  n_\text{ChEFT} = 1.3\, n_0$,  the $c_s^2$ posterior bands increase more rapidly compared to the ChEFT constraint which remains at softer sound velocities.  Accordingly,  choosing $n_\text{ChEFT} = 2.0\,n_0$ has a huge impact on the description of neutron stars: the posterior credible bands in Fig.\,\,\ref{fig:PosteriorBandsBW} become more tight and the stiffening of the speed of sound is delayed to energy densities $\varepsilon \gtrsim 300\,$MeV$\,$fm$^{-3}$.  With this change the central density of a $2.1\,M_\odot$ neutron star is increased to $n_c = 3.9_{-0.8}^{+0.6}\,n_0$. The softening seen in the ChEFT results around $n\sim 2\,n_0$ is in opposition to the apparent trend inferred from current astrophysical data. This slight tension was already noted in Ref.\,\,\cite{Essick2020}.  There it was suggested that the range of ChEFT applicability be left as a free parameter in the range $n_\text{ChEFT} \sim 1.1 - 2.0\, n_0$,  to be sampled together with the other parameters of the EoS and then marginalised in the end \cite{Han2023}. 

It thus appears that the EoS resulting from ChEFT at densities around twice $n_0$ has a tendency of becoming too soft in comparison with the conditions provided by astrophysical data.  Heavy ion collisions at intermediate energies probing the density region $n\sim 2-3\,n_0$ may help to further clarify this situation \cite{Kumar2023,Sorensen2023}.  An analysis of data from the FOPI experiment led to pressure constraints similar to those derived from ChEFT \cite{Huth2022}.  On the other hand,  the PREX II measurement of the $^{208}$Pb neutron skin thickness suggests a stiffer EoS \cite{Reed2021,Lim2022}.  Further insights can be expected from continuing  developments in the near future.

\subsection{Impact of asymptotic (pQCD) constraint}
\label{sec:ImpactpQCD}

\begin{figure}[tbp]
	\begin{center}
		\includegraphics[height=55mm,angle=-00]{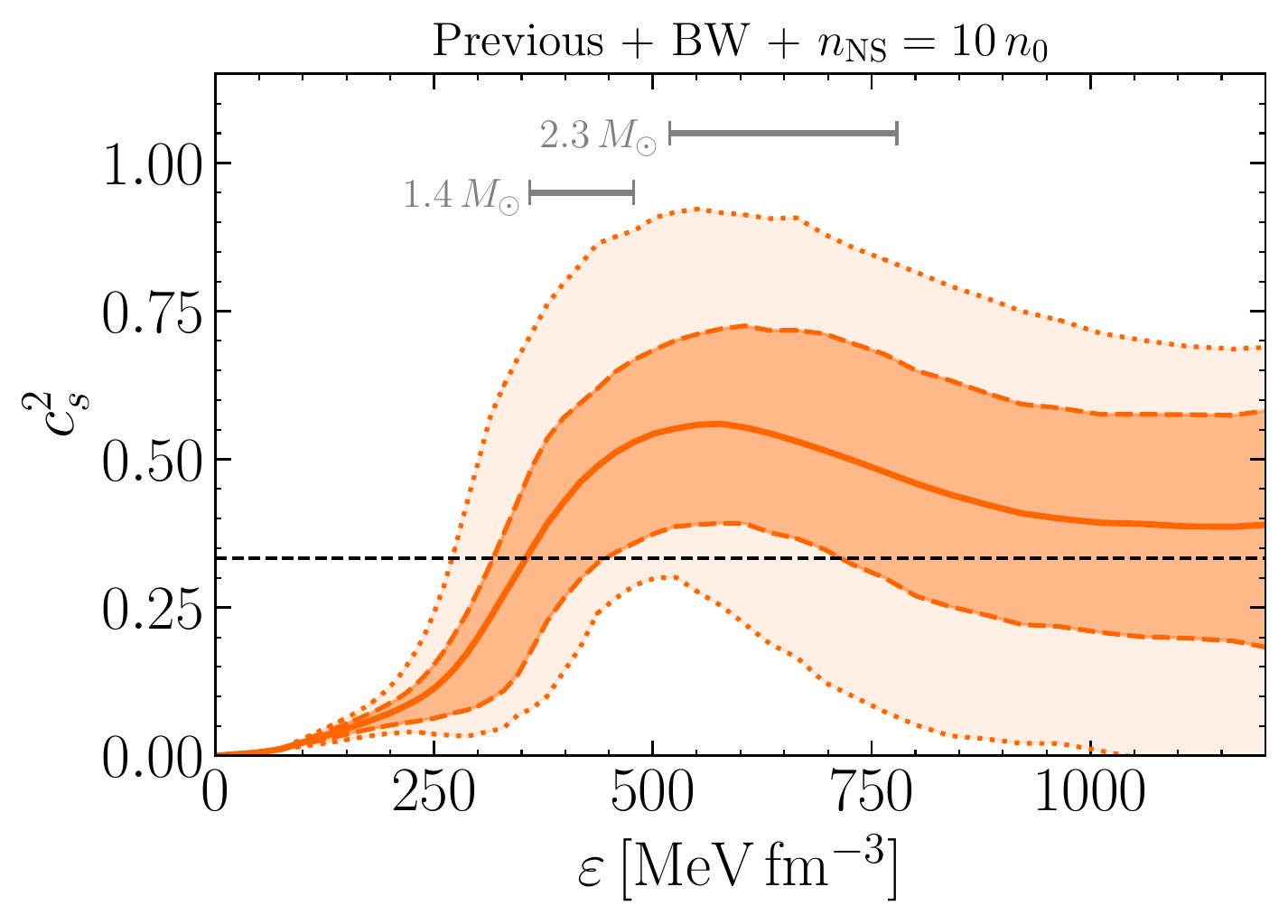}
		\caption{Similar to Fig.\,\,\ref{fig:PosteriorBandsBW}: posterior credible bands for the squared speed of sound,  $c_s^2$,  as a function of energy density $\varepsilon$,  based on the previous data as well as the new information from the black-widow pulsar (BW).  Here, however, the integral pQCD likelihood is implemented, as in Refs.\,\,\cite{Komoltsev2022,Gorda2023,Annala2023},  at $n_\text{NS} = 10\,n_0$ instead of $\varepsilon_\text{NS} = \varepsilon_{c,\max}$. }
		\label{fig:PosteriorBandspQCD}
	\end{center}
\end{figure}

Fig.\,\,\ref{fig:PosteriorBandspQCD} shows the posterior credible bands of $c_s^2(\varepsilon)$ using a different implementation of the asymptotic pQCD constraint:  we shift the matching condition at which it is verified whether the asymptotic pQCD requirement can be met in a causal and thermodynamically stable fashion,  from $\varepsilon_\text{NS} = \varepsilon_{c,\max}$ to a fixed point, $n_\text{NS} = 10\,n_0$. This is how the pQCD likelihood is implemented in Refs.\,\,\cite{Komoltsev2022,Gorda2023,Annala2023}.  It is obvious that this choice leads to strong,  even qualitative changes in the speed-of-sound credible bands.  While at energy densities $\varepsilon \lesssim 500\,$MeV$\,$fm$^{-3}$,  the posterior credible bands look similar to those in Fig.\,\,\ref{fig:PosteriorBandsBW},  at higher energy densities the speed of sound falls off and reaches significantly smaller values of $c_s^2$. This test case behaviour agrees well with the findings in Refs.\,\,\cite{Komoltsev2022,Gorda2023,Annala2023}.  The differences resulting from the two pQCD implementations has already been pointed out in Refs.\,\,\cite{Somasundaram2023a,Essick2023}.  

If the integral pQCD likelihood of Eq.\,(\ref{eq:LikelihoodpQCD}) is imposed at densities as high as $n_\text{NS} = 10\,n_0$,  EoSs featuring large sound speeds at energy densities $\varepsilon < 1200\,$MeV$\,$fm$^{-3}$ are in fact not excluded,  but they become much less likely.  In order to fulfil the integral pQCD constraint at higher densities,  EoSs with large sound speeds must decrease to smaller $c_s^2$,  whereas small sound speeds have more freedom to gain support in the analysis.  In fact,  imposing the integral pQCD likelihood at $n_\text{NS} = 10\,n_0$ makes large speeds of sound unlikely all the way down to $\varepsilon \sim 0$.  This softening does, however,  depend sensitively on the specific choice of $n_\text{NS}$.  For example,  an alternative scenario with $n_\text{NS} = 8\,n_0$ which is also far beyond the central densities of most neutron stars,  leads to a much less pronounced softening in $c_s^2$.  Despite these drastic changes,  in our analysis as in Refs.\,\cite{Gorda2023,Ecker2022},  the mass-radius relation is only weakly affected.  Small modifications are seen only for the most massive neutron stars, $M > 2.1\,M_\odot$, which are no longer constrained by radius measurements.  In fact,  even the properties of a $2.3\,M_\odot$ neutron star depicted in Tab.\,\,\ref{tab:NS_properties3} change only slightly.  The strong evidence against small sound speeds in the cores of neutron stars with masses $M \leq 2.0\,M_\odot$ persists.

Because the EoSs beyond $\varepsilon_{c,\max}$ are no longer constrained by astrophysical data but merely interpolated up to high densities,  we believe that $\varepsilon_\text{NS} = \varepsilon_{c,\max}$ is the better (i.e. more conservative) choice. Selecting a higher matching density may lead to an overestimation of the pQCD impact. With such a higher matching density beyond the range of control by data the impact is expected to depend sensitively on the choice of priors in the unconstrained interpolation region \cite{Essick2023}.  With the conservative choice, employed in the present work,  however,  the integral pQCD likelihood of Eq.\,(\ref{eq:LikelihoodpQCD}) has a negligible influence on the sound speed and related properties of neutron stars.  This corresponds well to the conclusions drawn in Ref.\,\,\cite{Somasundaram2023a} where the authors also find only a very small impact of the pQCD integral constraint imposed at $\varepsilon_{c,\max}$.

Recently some authors have claimed evidence for a possible phase transition to a new state of matter near the maximum neutron star mass, $M_{\max}$ \cite{Annala2023,Han2023},  based on inferred values for the sound speed \cite{Ecker2022,Jiang2023},  the polytrope index \cite{Annala2020} or the behaviour of the trace anomaly measure at densities corresponding to $M_{\max}$.  In these analyses the authors use the approach with fixed matching density to implement the asymptotic pQCD constraint,  which,  as we discussed,  is strongly prior dependent. In addition the central density reached inside the most massive neutron stars are, similar to our analysis, much higher than the central density of a $2.1\,M_\odot$ neutron star. This high density regime is, however, only loosely constrained by the current astrophysical data. Therefore, analyses of properties of the most massive stars should require a particularly detailed assessment of the prior dependence induced by the interpolation to high densities. In contrast, in our analysis we claim evidence only for neutron stars with masses $\lesssim 2.1\,M_\odot$, which are still in the density regime that is well-constrained by the current astrophysical data.

A further note concerns selected recent analyses which see a pronounced softening of the sound velocity and equation of state at high energy densities \cite{Annala2023,Han2023,Marczenko2023,Jiang2023,Altiparmak2022}.  Apart from a different implementation of the pQCD likelihood,  as discussed above,  this may be due to differences in presentation.  As pointed out,  in computing credibility bands we follow Refs.\,\,\cite{Greif2019,Raaijmakers2021} and employ each equation of state only up to their respective maximum central energy density $\varepsilon_{c,\max}$ corresponding to the respective maximum mass.  Similarly,  in the computation of the credible bands for the radius as a function of mass we consider each EoS only up to their respective maximum mass $M_{\max}$,  as higher masses do not correspond to stable neutron stars.  In contrast,  Refs.\,\,\cite{Annala2023,Han2023,Marczenko2023,Jiang2023,Altiparmak2022} among others use each EoS up to arbitrarily high energy densities.  We decided against this approach  because it generates an uncontrolled mix of information from EoSs constrained by astrophysical data and EoSs beyond empirical limits.  If we also use instead each equation of state up to arbitrary $\varepsilon$,  we do indeed observe a slight softening in Fig.\,\,\ref{fig:PosteriorBandsBW}.  However,  this behaviour is not based on the empirical data but merely on the interpolation extending up to the pQCD constraint at asymptotic densities.

\subsection{Possible impact of HESS J1731-347}

\begin{figure*}[tbp]
	\begin{center}
		\includegraphics[height=55mm,angle=-00]{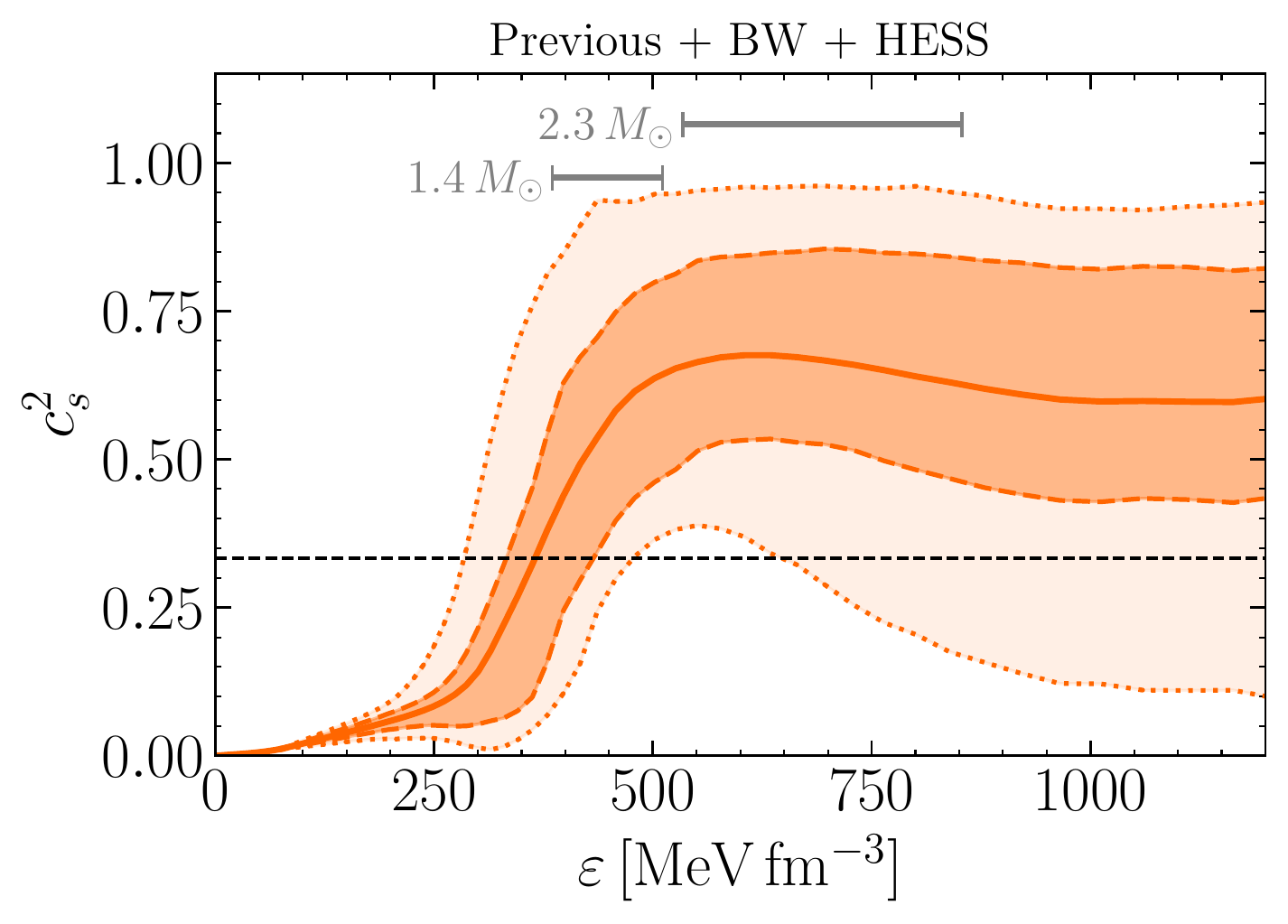} 
		\includegraphics[height=55mm,angle=-00]{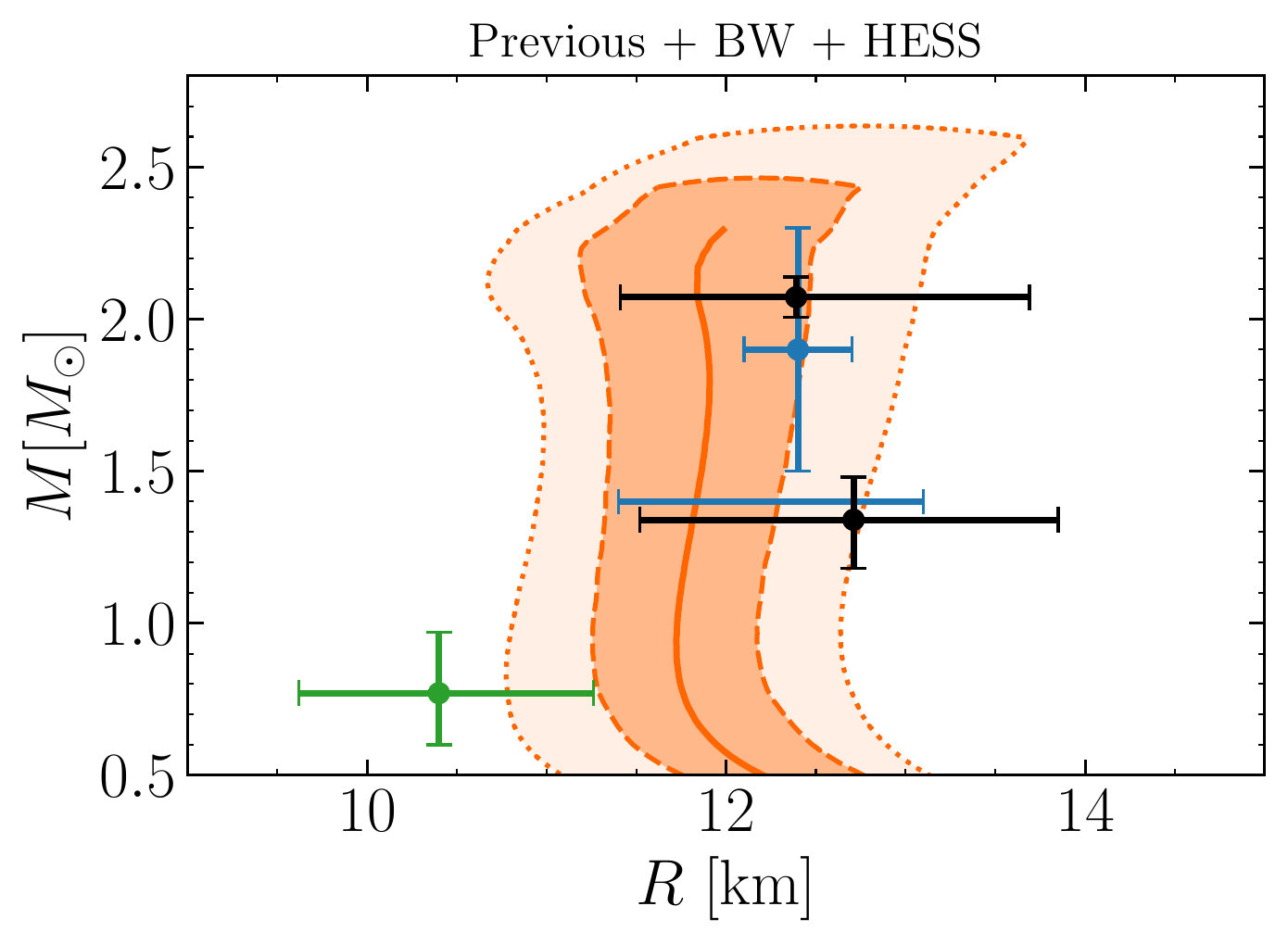} \\
		\caption{Similar to Fig.\,\,\ref{fig:PosteriorBandsBW}: posterior credible bands for the squared speed of sound, $c_s^2$,  as a function of energy density $\varepsilon$,  and mass-radius relation 
$R(M)$,  but now including the information from the supernova remnant HESS J1731-34 in addition to the black-widow pulsar PSR J0952-0607. The resulting mass-radius relation is also compared to the marginalized intervals at the 68\% level from the analysis of the low-mass supernova remnant HESS J1731-34 \cite{Doroshenko2022} (green).}
		\label{fig:PosteriorBandsHESS}
	\end{center}
\end{figure*}

Finally,  we analyse the impact of including the mass-radius estimate for the very light central compact object HESS J1731-347 reported in Ref.\,\,\cite{Doroshenko2022}.  The updated posterior credible bands including this additional information are collected in Fig.\,\,\ref{fig:PosteriorBandsHESS}. To reach small radii, the speed of sound has to increase more rapidly at densities above the ChEFT constraint at $n_\text{ChEFT} = 1.3\,n_0$. With the new information,  the credible bands are much more tightly constrained compared to Fig.\,\,\ref{fig:PosteriorBandsBW}. The inclusion of the supernova remnant in the Bayesian analysis shifts the radii at all masses to lower values. The radius of a $1.4\,M_\odot$ neutron star reduces to $R=11.8_{-0.4}^{+0.5}\,$km at the 68\% level,  similar to the value $R=11.7\pm0.5\,$km reported in Ref.\,\,\cite{Doroshenko2022} at the 90\% level.  In addition to HESS J1731-347 and the previously available data listed in Tab.\,\,\ref{tab:DataSet},  the latter estimate includes additional information from the x-ray burster 4U 1702-429 and from the rotation limit for the radio pulsar PSR J1748-2446ad. There is visibly some tension between the radius estimate at lower masses based on the current data,  most importantly from PSR J0030+0451,  and HESS J1731-347,  as already noted in Ref.\,\,\cite{Jiang2023}.  As a consequence of this tension the posterior credible band for $R(M)$ agrees with the credible intervals of the supernova remnant only marginally at the 95\% level.  However,  as also noted in Sec.\,\,\ref{sec:LikelihoodHESS}, the analysis of HESS J1731-347 involves more systematic uncertainties compared to other data including NICER.

In Tab.\,\,\ref{tab:NS_properties3} we show inferred properties of neutron stars with mass $M = 0.77\,M_\odot$. The central density of a $0.77\,M_\odot$ neutron star is low,  only $n_c = 2.2\pm0.3\,n_0$.  As an additional output,  the Bayes factor $\mathcal{B}_{\Delta \geq 0}^{\Delta < 0}$ with the inclusion of the HESS supernova remnant increases from about 8 to $\mathcal{B}_{\Delta \geq 0}^{\Delta < 0} \simeq 11$,  so that the evidence for a negative trace anomaly measure inside neutron stars turns from moderate to strong. 

\setlength{\extrarowheight}{6pt}
\begin{table}[tbp]
	\centering
	\begin{tabularx}{\linewidth}{|l|l|lXX|}
		\hline \hline 
		\multicolumn{2}{|l|}{} & \multicolumn{3}{l|}{\hspace*{1mm}Previous + BW + HESS} \\
		\multicolumn{2}{|l|}{} && 95\% & 68 \%   \\ \hline
		&$n_c /n_0$ && $2.2_{-0.5}^{+0.6}$ & $\pm 0.3$  \\
		&$\varepsilon_c  \, $[MeV$\,$fm$^{-3}$] && $352_{-87}^{+81}$ & $_{-52}^{+38}$  \\
		$0.77\, M_\odot$& $P_c \, $[MeV$\,$fm$^{-3}$] && $25_{-7}^{+9}$ & $\pm 4$ \\
		&$R \, $[km] && $11.8_{-1.0}^{+0.9}$ & $_{-0.4}^{+0.5}$ \\
		&$\Lambda$ && $7911_{-3352}^{+3979}$ & $_{-1827}^{+1799}$  \\ [.8ex]
		\hline \hline 
	\end{tabularx}
	\caption{Same as Tab.\,\,\ref{tab:NS_properties1}, but median and credible intervals for a neutron star with mass $M = 0.77 \, M_\odot$ are displayed,  given the previously available data, the mass measurement of the black widow pulsar and the mass-radius data of the supernova remnant HESS J1731-347.}
	\label{tab:NS_properties3}
\end{table}
\setlength{\extrarowheight}{4pt}

\section{SUMMARY AND CONCLUSIONS}
\label{sec:Summary}

The present work is a substantial update beyond our previous Bayesian studies focusing on the equation-of-state of neutron star matter.  A primary aim is to tighten the conclusions about possible phase transitions in the cores of neutron stars.  The framework has been expanded in several respects.  New information from the mass determination of the heavy black-widow pulsar PSR J0952-0607 has been included in the inference procedure, in addition to the previously established empirical data base from NICER,  Shapiro delay measurements and binary neuron star merger observations.  A novel likelihood implementation of the low-density constraint from ChEFT has been introduced. A detailed assessment of the matching condition connecting the neutron star region with the perturbative QCD limit at asymptotic densities has been performed.  Finally,  the influence of the unusually light supernova remnant HESS J1731-347 on the overall systematics of neutron star properties has been examined.

The results and conclusions are summarized as follows:
 \begin{enumerate}
	\item The low-density,  nuclear physics constraint from ChEFT is treated as a likelihood rather than a prior,  and its conservative application is restricted to baryon densities $n \leq 1.3\,n_0$ (with $n_0$  the density of equilibrium nuclear matter).  In the range $\varepsilon \lesssim 0.5\,$GeV$\,$fm$^{-3}$ the squared speed of sound,  $c_s^2 = \partial P/\partial\varepsilon$,  rises rapidly beyond $c_s^2 = 1/3$  and develops a plateau at larger energy densities.  Moderate tension exists between ChEFT extrapolations of $c_s^2$ up to $n \simeq 2\,n_0$ (where $n_0 = 0.16\,$fm$^{-3}$) and the trend towards a stiffer EoS implied by the astrophysical data.  
	\item A Bayes factor analysis demonstrates extreme evidence that the sound velocity exceeds the conformal bound,  $c_s \leq \sqrt{1/3}$,  in neutron stars. 
	\item The incorporation of the heaviest neutron star observed so far,  PSR J0952-0607,  with rotation corrections properly applied,  results in a further stiffening of the EoS, $P(\varepsilon)$.  The increased pressure covers the entire range of energy densities realized in the cores of neutron stars with masses M = $1.4 - 2.3\, M_\odot$.  The TOV maximum supported mass is raised to $M_\text{max} = 2.31^{+0.14}_{-0.17}\,M_\odot$.  
	\item As a consequence of the stiffer EoS,  the central core densities of neutron stars are reduced: to $n_c < 3\,n_0$ for $M = 1.4\,M_\odot$ and to $n_c < 5\,n_0$ even for masses as high as $M = 2.3\,M_\odot$ (at 68\% credibility).  This observation is of some significance for a possible interpretation of neutron star matter in terms of baryonic degrees of freedom: the average distance between two baryons in the core of a 1.4 $M_\odot$ star is still about 1.2$\,$fm.  Even in the core centre of a 2.3 $M_\odot$ neutron star,  this mean distance does not fall below about 1$\,$fm, well beyond the characteristic hard-core radius $r\sim 0.5\,$fm of the short-range repulsive interaction that keeps two nucleons apart. 
	\item Within the posterior credible bands (at 68\% level),  possible phase coexistence regions,  i.e. domains of constant pressure in a Maxwell construction,  are restricted to a maximal width of $\Delta n/n \leq (\Delta n/n)_{\max} \simeq 0.2$.  This upper limit stays nearly constant throughout the density regime relevant to neutron stars,  allowing at most for a weakly first-order phase transition which would have little observable impact. 
	\item The inclusion of the heavy black-widow pulsar increases the evidence against small sound speeds inside neutron star cores.  A corresponding Bayes factor investigation demonstrates strong evidence against minimum squared sound speeds smaller than $c_{s,\min}^2 \leq 0.1$,  indicative of a possible strong first-order phase transition,  in neutron stars with masses up to $M \leq 2.1\,M_\odot$.  This is consistent with other studies which find $c_{s}^2 > 0.1$ at the 95\% level in the cores of neutron stars with mass $M \simeq 2.0\,M_\odot$.
	\item Extreme evidence is established on the basis of corresponding Bayes factors against scenarios with multiple stable mass-radius branches, including twin-star solutions of the TOV equations.  The low-density constraint from ChEFT plays a crucial role in drawing this conclusion.  In the absence of this constraint the evidence against twin-star scenarios turns out to be `only' strong.
	\item Matching the neutron star EoS to asymptotic pQCD requires an extrapolation from densities reached in neutron star cores, $n_\text{NS}$,  to the extreme densities at which perturbative QCD methods can be applied.  If instead a matching density is chosen at a value far beyond the density range controlled by empirical observations,  the impact of the pQCD constraint on the EoS for neutron stars depends sensitively on this prior choice. As a result this pQCD impact may be overestimated.  With the matching density $n_\text{NS}$ fixed at the value corresponding to the maximum supported neutron star mass for each EoS,  we find only very little impact of the asymptotic pQCD.
	\item The trace anomaly measure,  $\Delta = 1/3 - P/\varepsilon$, is a presently much discussed quantity that provides an estimate for the approach to conformal matter at high baryon densities.  A corresponding Bayes factor analysis suggests moderate evidence for a negative trace anomaly in heavy neutron stars ($M \gtrsim 2\,M_\odot$).  This implies that $\Delta$ should change its sign to positive at some higher densities in order to approach the asymptotic limit,  $\Delta\rightarrow 0$,  from above. 
	\item The mass-radius relation for neutron stars inferred from previous plus new heavy-mass data has the remarkable feature that the median is at an almost constant radius ($R\simeq 12.3 $ km) for all masses above $M\gtrsim0.7\,M_\odot$.  The ultralight supernova remnant HESS J1731-347,  if incorporated into the inference procedure,  falls out of this systematics and would shift this mass-radius relation to smaller radii. In particular, the inferred HESS radius including uncertainties is located outside the 68\% credible band of the mass-radius relation.  
\end{enumerate} 

We thus come to the overall conclusion that the additional incorporation of the new massive black-widow pulsar data in the Bayesian inference analysis performed with very large ($\sim 10^6$) EoS samples further strengthens the evidence against very low squared sound velocities in neutron stars. This together with a maximum possible phase coexistence region of $(\Delta n/n)_{\max} \simeq 0.2$ within the 68\% posterior credible bands renders the occurrence of a strong first-order phase transition in neutron star cores more unlikely.  On the other hand,  a continuous hadrons-to-quarks crossover or a conventional baryonic matter scenario are not ruled out.  We are looking forward towards progressively more rigorous constraints on the EoS as the quantity and quality of observational data increases in the future.

\begin{acknowledgments}
	Many thanks go to Kenji Fukushima,  Oleg Komoltsev,  Michal Marczenko,  Larry McLerran and Krzysztof Redlich for stimulating discussions and communications.  This work has been supported in part by Deutsche Forschungsgemeinschaft (DFG) and National Natural Science Foundation of China (NSFC) through funds provided by the Sino-German CRC110 “Symmetries and the Emergence of Structure in QCD” (DFG Grant No. TRR110 and NSFC Grant No. 11621131001),  and by the DFG Excellence Cluster ORIGINS.
\end{acknowledgments}

\appendix
\section{ROTATION CORRECTION}
\label{sec:RotationAdjustment}

\begin{figure}[tbp]
	\begin{center}
		\includegraphics[height=55mm,angle=-00]{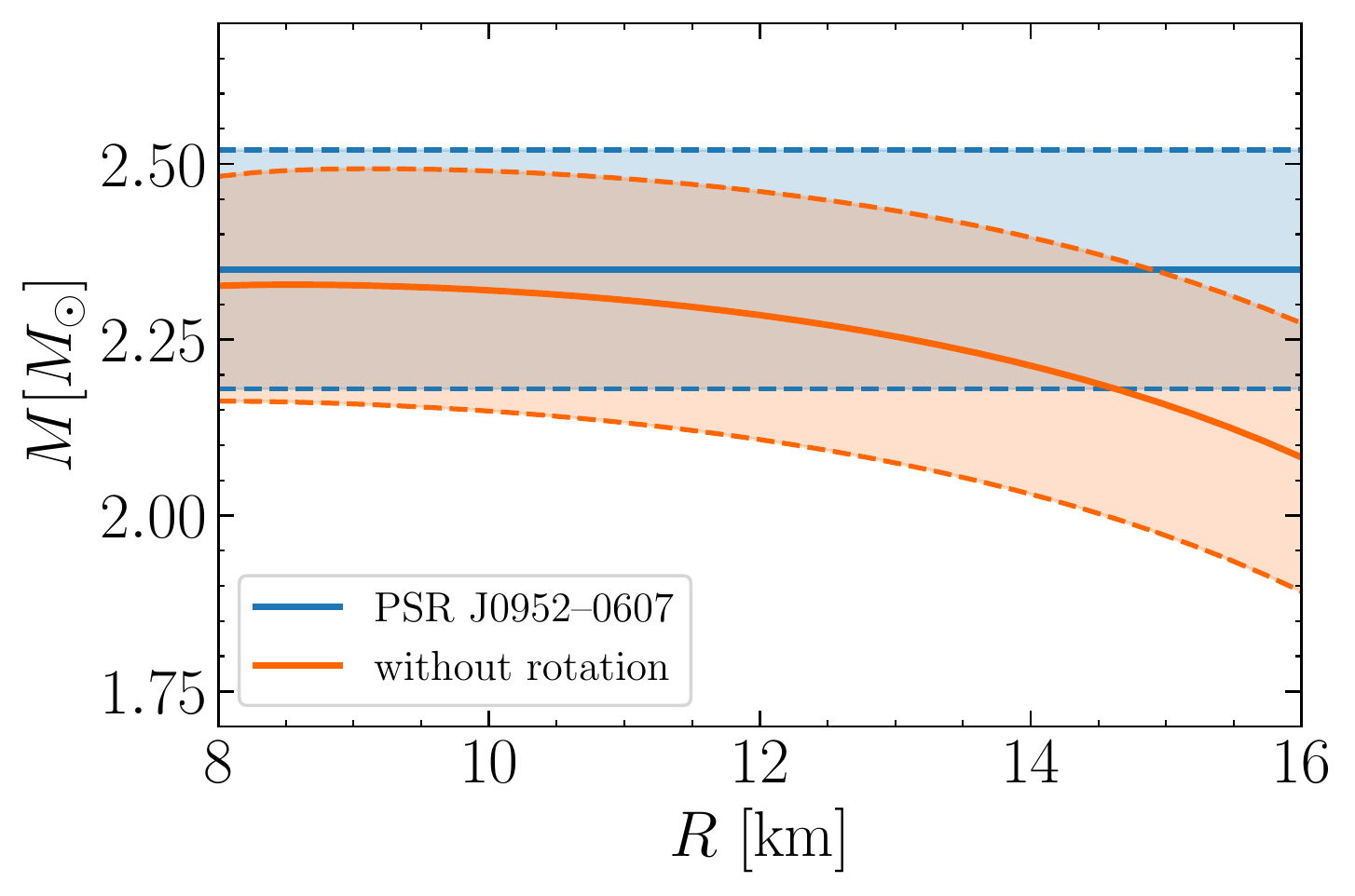}
		\caption{PSR J0952-0607 mass measurement with 68\% uncertainty \cite{Romani2022} compared to the static case (without the $\nu \simeq 709\,$Hz rotation) computed using the prescription in Ref.\,\,\cite{Konstantinou2022}.}
		\label{fig:RotationCorrection}
	\end{center}
\end{figure}

The magnitude of the relative mass increase of the black widow pulsar PSR J0952-0607 by its high spin frequency of $\nu \simeq 709\,$Hz is illustrated in Fig.\,\,\ref{fig:RotationCorrection}.  The correction is estimated using the EoS-independent formulae deduced in Ref.\,\,\cite{Konstantinou2022}.  This adjustment is quite strongly radius-dependent,  with a significant difference between rotating and non-rotating mass at larger radii.  The uncertainty in the correction as reported in Ref.\,\,\cite{Konstantinou2022} is smaller than the uncertainty in the heavy-mass measurement itself.  At a neutron star radius of $R = 12\,$km the rotating mass $M = 2.35\,M_\odot$ decreases by 3\% to an equivalent non-rotating mass of $M\simeq 2.28\,M_\odot$.

\section{CHEMICAL POTENTIAL \&\\ EOS TABLE}
\label{sec:ThermoDynQuantities}

\begin{figure}[tbp]
	\begin{center}
		\includegraphics[height=55mm,angle=-00]{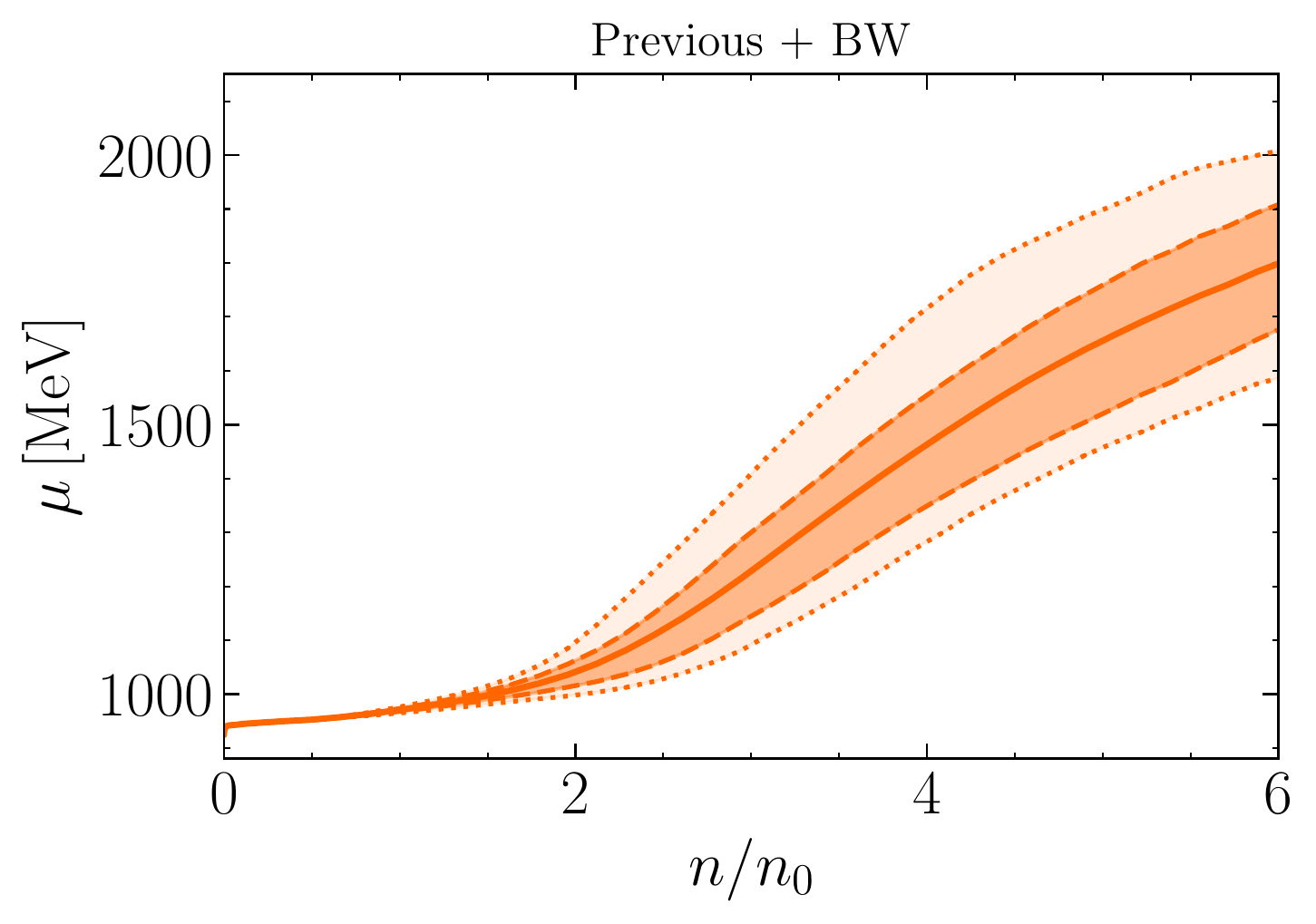} \\
		\caption{Posterior 95\% and 68\% credible bands and medians for the baryon chemical potential $\mu$ as a function of baryon density $n$ in units of the nuclear saturation density,  $n_0 = 0.16\,$fm$^{-3}$.}
		\label{fig:Mu}
	\end{center}
\end{figure}

For a given equation of state the baryon chemical potential can be computed as
\begin{equation}
	\mu = {\partial\varepsilon\over\partial n} = \frac{\varepsilon + P}{n} ~.
	\label{eq:mu}
\end{equation}
The resulting posterior credible bands are displayed in Fig.\,\,\ref{fig:Mu}.  Note that this baryon chemical potential corresponds to the total $\mu$ from all active degrees of freedom carrying baryon number: $\mu = \sum_i x_i \mu_i$, where $x_i = n_i/n$ is the fraction corresponding to each species $i$. Our agnostic approach for the speed of sound does not permit to distinguish between separate species of constituents realized inside neutron stars.

For practical purposes and applications,  the median values of the squared sound velocity $c_s^2(\varepsilon)$ as a function of energy density,  as shown in Fig.\,\,\ref{fig:PosteriorBandsBW},  are listed in Tab. \ref{tab:ThermoDynQuantities},  using the data set including the previously available data and the new information from PSR J0952-0607.  Based on these values the pressure $P$ is computed using Eq.\,(\ref{eq:soundspeed}),  as well as the baryon density $n$ and the chemical potential $\mu$ with Eq.\,(\ref{eq:mu}).  The asymmetry of the posterior distribution causes small insignificant deviations between the pressure computed from the integral of the sound velocity and the median of $P$ in Fig.\,\,\ref{fig:PosteriorBandsBW}.  

A comparison of this table with the corresponding table in Appendix C of Ref.\,\cite{Brandes2023} is instructive as it underlines the significant stiffening of the EoS that emerges when the new PSR J0952-0607 data are incorporated  together with the updated implementations of ChEFT and pQCD constraints.  For example,  in the range $\varepsilon = 0.5-0.9\,$GeV$\,$fm$^{-3}$ the pressure has increased by typically about one third as compared to the previous results in Ref.\,\cite{Brandes2023}.

\begin{table}[tbp]
	\centering
	\begin{tabularx}{\linewidth}{|l|X|l|X|X|}
		\hline \hline  
		$\varepsilon\,[$GeV\,fm$^{-3}]$ & $c_s^2$ & 
		$P\,[$MeV\,fm$^{-3}]$ & $n/n_0$ & $\mu\,$[GeV] \\ \hline
		0.1 & 0.02 & 0.8 & 0.66 & 0.96 \\ 
		0.2 & 0.07 & 5.4 & 1.31 & 0.98 \\ 
		0.3 & 0.20 & 17.4 & 1.93 & 1.03 \\ 
		0.4 & 0.44 & 49.0 & 2.50 & 1.12 \\ 
		0.5 & 0.59 & 101.3 & 3.03 & 1.24 \\ 
		0.6 & 0.64 & 163.0 & 3.52 & 1.35 \\ 
		0.7 & 0.64 & 226.8 & 3.97 & 1.46 \\ 
		0.8 & 0.62 & 289.8 & 4.39 & 1.55 \\ 
		0.9 & 0.60 & 350.9 & 4.78 & 1.64 \\ 
		1.0 & 0.59 & 410.6 & 5.15 & 1.71 \\ 
		1.1 & 0.60 & 470.4 & 5.50 & 1.78 \\ 
		1.2 & 0.61 & 531.1 & 5.84 & 1.85 \\  
		\hline \hline 
	\end{tabularx}
	\caption{Tabulated values of the median for the squared sound velocity,  $c_s^2$,  as a function of energy density $\varepsilon$ as shown in Fig.\,\,\ref{fig:PosteriorBandsBW}, i.e. including the previously available data and the new information from PSR J0952-0607. Based on these values the pressure is computed as well as the baryon density $n$ (in units of the nuclear saturation density $n_0$) and the baryon chemical potential $\mu$.}
	\label{tab:ThermoDynQuantities}
\end{table}

\section{BAYES FACTOR TABULAR}
\label{sec:BayesFactorTable}

A key result of the present work is the systematics of the Bayes factor $\mathcal{B}^{c_{s,\min}^2 > 0.1}_{c_{s,\min}^2 \leq 0.1}$,  quantifying the evidence against a dropping of the squared sound speed to values below 
$c_{s,\min}^2 \leq 0.1$,  as a function of the maximum mass in neutron stars.  For a further documentation of these results a table of numerical values is useful to underscore the increase of this evidence when the new information from PSR J0952-0607 is incorporated.

There is extreme evidence that the minimum squared sound speed,  after exceeding the conformal limit,  does not drop to values smaller than $0.1$ for neutron stars with masses $M \leq 1.9 \, M_\odot$.  There is strong evidence that $c_{s,\min}^2$ does not become smaller than $0.1$ in neutron stars with mass $M \leq 2.0\, M_\odot$. The Bayes factors increase further with the inclusion of the black widow (BW) mass data.  With this new empirical information  there is strong evidence against small sound speeds $c_{s,\min}^2 <0.1$ inside neutron stars even up to masses $M \leq 2.1 \, M_\odot$.

\begin{table}[tbp]
	\centering
	\begin{tabularx}{\linewidth}{|X|lXX|}
		\hline \hline
		&&& \\ [-3ex]
		& \hspace{1mm}& \multicolumn{2}{l|}{$\mathcal{B}^{c_{s,\min}^2 > 0.1}_{c_{s,\min}^2 \leq 0.1}$} \\ [.8ex]
		$M /M_\odot$ && Previous & Previous + BW \\ \hline
		1.9 && 201.02 & 500.86 \\
		2.0 && 46.26 & 229.80 \\
		2.1 && 4.55 & 15.00 \\
		2.2 && 1.88 & 3.63 \\
		2.3 && 1.45 & 2.16 \\
		\hline \hline  
	\end{tabularx}
	\caption{Bayes factors $\mathcal{B}^{c_{s,\min}^2 > 0.1}_{c_{s,\min}^2 \leq 0.1}$ comparing EoS samples with the following competing scenarios: a) minimum squared speed of sound  (following a maximum),  with $c^2_{s,\min}$ larger than 0.1,  excluding a strong first-order phase transition with a Maxwell construction; versus b) EoS samples with $c_{s,\min}^2 \leq 0.1$.  The Bayes factors are calculated for a given maximum neutron star mass $M$,  i.e.  the minimum speed of sound up to the corresponding maximum mass is used. in the computation. }
	\label{tab:BayesFactorSmallCsM}
\end{table}

\bibliography{new_measurements_library}

\end{document}